\documentclass[aps,amsmath,amssym,amsfonts,floats,floatfix,twocolumn]{revtex4-1}

\usepackage{graphicx}
\usepackage{float}
\usepackage[usenames,dvipsnames]{color}
\newcommand{\sech}{\mathrm{sech}}

\newcommand{\ev}{{\bf \hat{e}}}
\newcommand{\uv}{{\bf u}}
\newcommand{\xv}{{\bf x}}
\newcommand{\rv}{{\bf r}}

\begin{document}

%\title[Short title]{Self-assembling ribbons of trapezoidal particles with cohesive and repulsive interactions self-limit at any target width}

\title{Stress accumulation versus shape flattening in frustrated, warped-jigsaw particle assemblies}
\author{Isaac R. Spivack$^{1,2}$, Douglas M. Hall$^1$ and Gregory M. Grason$^1$}
\address{$^1$ Department of Polymer Science and Engineering, University of Massachusetts-Amherst}
\address{$^2$ Department of Physics, University of Michigan}

\begin{abstract}
Geometrically frustrated assembly has emerged as an attractive paradigm for understanding and  engineering assemblies with self-limiting, finite equilibrium dimensions.  We propose and study a novel 2D particle based on a so-called ``warped jigsaw'' (WJ) shape design:  directional bonds in a tapered particle favor curvature along multi-particle rows that frustrate 2D lattice order.  We investigate how large-scale intra-assembly stress gradients emerge from the microscopic properties of the particles using a combination of numerical simulation and continuum elasticity. WJ particles can favor anisotropic ribbon assemblies, whose lateral width may be self-limiting depending on the relative strength of cohesive to elastic forces in the assembly, which we show to be controlled by the range of interactions and degree of shape misfit. The upper limits of self-limited size are controlled by the crossover between two elastic modes in assembly: the accumulation of shear with increasing width at small widths giving way to unbending of preferred row curvature, permitting assembly to grow to unlimited sizes. We show that the stiffness controlling distinct elastic modes is governed by combination and placement of repulsive and attractive binding regions, providing a means to extend the range of accumulating stress to sizes that are far in excess of the single particle size, which we corroborate via numerical studies of discrete particles of variable interactions. Lastly, we relate the ground-state energetics of the model to lower and upper limits on equilibrium assembly size control set by the fluctuations of width along the ribbon boundary.

\end{abstract}

\maketitle

\section{Introduction}

Geometric frustration is a broadly known phenomenon in condensed matter systems, defined as the impossibility of perfectly satisfying  local interactions globally throughout a system~\cite{Sadoc2006}. Historically, the concept was first associated with low-temperature states of magnetism and spin order~\cite{vannimenus_theory_1977}, but was subsequently generalized to complex states of soft matter~\cite{Kleman1989}, including blue phase liquid crystals~\cite{Sethna1983, Sadoc_2020} and amorphous sphere packings~\cite{Nelson1989}.  This classical view of frustration focuses on bulk states where geometric incompatibility is resolved by extensive array of topological defects~\cite{Sadoc2006, Kleman1989, Tarjus2005}, resulting in rough energy landscapes populated by extensive number of degenerate, or nearly degenerate, ground states.

\begin{figure*}
    \centering
    \includegraphics[width=0.9\linewidth,height=\textheight,keepaspectratio]{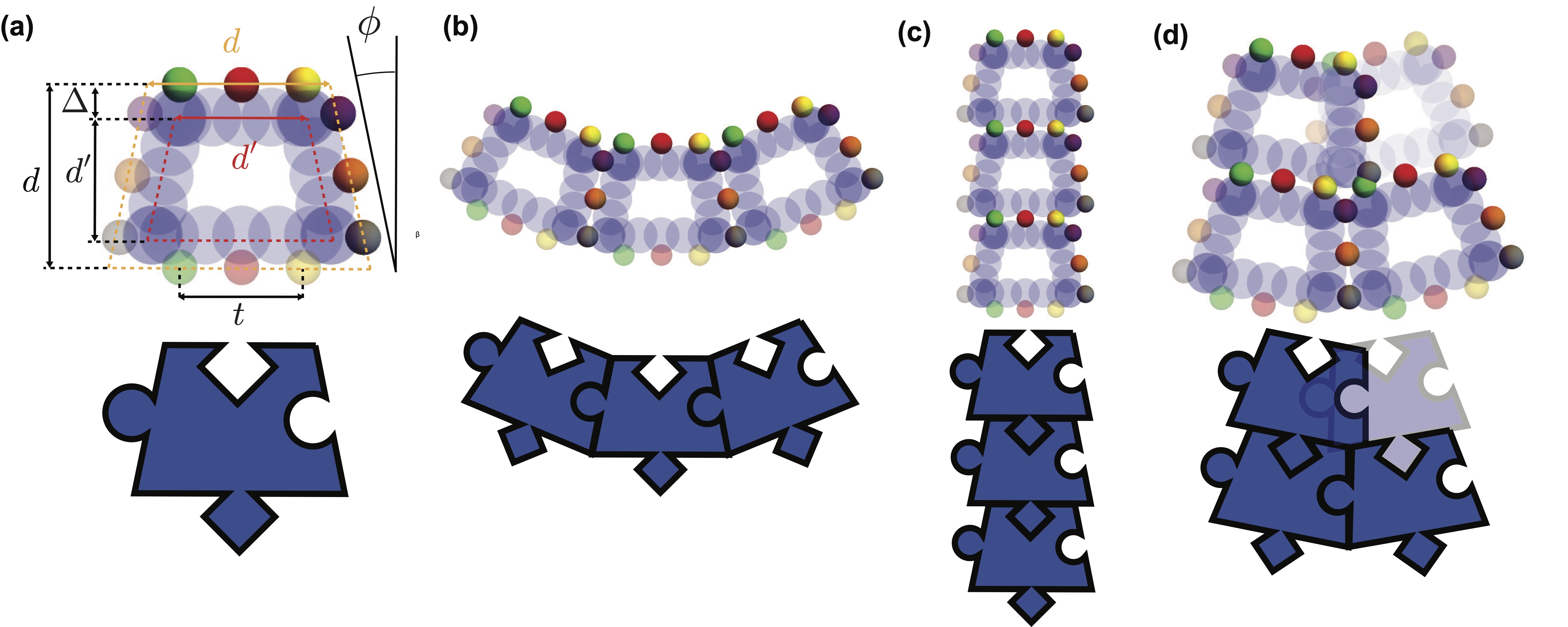}
    \caption{The warped-jigsaw particle model exhibits geometrically frustrated assembly.  (a) The particle consists of sites centered on two nested isosceles trapezoids, with spacing $\Delta$ between them. The respective heights and smaller bases are equal, with values $d$ and $d'$, where $d$ is the nominal particle spacing. The angle $\phi$ controls the amount of frustration, so that the unfrustrated case $\phi=0$ is a square particle shape. Along each side, the attractor spacing is $t$, with repulsive sites at the centers of each side and halfway between attractors. The blue spheres shown with centers on the inner trapezoid are volume exclusion sites, while the smaller colored spheres on the outer trapezoid are attractive and repulsive sites. The interactions are specific: for example, the red repulsive site on the top of a WJ particle only interacts with the opaque red site on the bottom of other particles. This leads to a lock-and-key mechanism as in the puzzle piece representation of the particles shown in the lower row.  (b-c) The shape of the particles causes them to curl up into rows in the $x$-direction and form straight lines in the $y$-direction. (d) Assembly of a 2D (planar) array is frustrated, as highlighted by the misfitting particle in the upper right.}
    \label{fig:1}
\end{figure*}

It has recently been recognized that frustration gives rise to new behaviors in self-assembling materials~\cite{Grason2016}, deriving from two key features.  First, constituent building blocks (e.g. polymers, colloids, proteins) are relatively ``soft" and held together by weak, non-covalent forces. Second, assemblies need not reach bulk states, and thus, have additional degrees of freedom associated with the size and shape of the assembled domain.  Unlike bulk or rigid systems where frustration must be resolved by defects, in soft assemblies shape-misfit can be tolerated, over at least some range of sizes, through the build up of smooth gradients in the subunit shapes and packings~\cite{Meiri2021}. The self organization of long-range intra-assembly stress gradients is a defining characteristic of {\it geometrically-frustrated assemblies} (GFAs), as it can give rise to scale-dependent thermodynamics without counterpart in canonical, unfrustrated assemblies.  A singular outcome of the scale-dependent ``misfit'' is {\it self-limiting assembly}~\cite{hagan_equilibrium_2020}. While the drive to maximize the number of cohesive contacts generically favors unlimited, bulk aggregates in assemblies, GFAs are subject to elastic penalties associated with the misfit gradients in the assembly. An example of this is illustrated schematically for the {\it warped jigsaw} particle model in ~figure \ref{fig:1}, which is the focus of this paper.  In GFA, the balance between cohesion and the superextensive costs of frustration can select equilibrium domain sizes that are finite and, in principle, arbitrarily larger than the subunits themselves.  That is, in contrast to the prevailing paradigms of geometric frustration in bulk systems, in GFAs the free boundaries themselves, which can be of variable shape and size, represent critical degrees of freedom.

This thermodynamic picture has been applied to understand anomalous assembly properties of a range of existing experimental soft matter systems, from chiral membranes~\cite{Ghafouri2005, Achard2005, Selinger2004, Aggeli2001, Achard2005, Zhang2019, Serafin2021}, crystalline domains on spherical surface~\cite{Schneider2005, Meng2014, Paquay17, Mendoza2020} to twisted fibers of filamentous proteins~\cite{Turner2003, Grason2007, Yang2010, Hall2016, Hall2017, Grason2020}, in part to rationalize observed finite domain formation.  Theoretical models in these cases rely almost exclusively on generalized continuum elastic models which parameterize the costs of cumulative gradients in local packing.  While these theories can provide a consistent thermodynamic model of these experimental systems, their predictive power is limited by the reliance on unknown phenomenological parameters, including the elastic moduli for different modes of assembly deformation and the effective ``strength'' of frustration.  

A recent interest has emerged to understand and design the accumulation of intra-aggregate stress and its thermodynamic stresses from the properties of the misfitting particles themselves.  That is, given an arbitrarily misfitting particle, what is the range of self-limiting morphologies it can exhibit and under what thermodynamic conditions?  One motivation to address this point was raised in a study by Lenz and Witten \cite{Lenz2017}, which pointed out that protein ``building blocks" tend to aggregate under a broad range of conditions, but don't necessarily take shape-complementary structures. As such, some degree of shape-frustration in multi-protein aggregation is likely the rule, rather than the exception, which motivated their study of the planar assembly of a select class of incompatible elastic polygons, which generically exhibited finite-width fiber formation upon simulated (non-equilibrium) assembly.  

A second motivation to address these questions comes from the desire to potentially engineer and {\it program} the finite size of equilibrium assemblies through the intentional frustration of their shapes.  An expanding array of synthetic techniques afford pathways to engineer self-assembling building blocks with controlled shapes and interactions~\cite{Glotzer2007,Su2020, Hueckel21}, from anisotropic colloids to shape programmed DNA and protein particles.  Attempting to capitalize on this potential, a recent study by Berengut {\it et al.} designed and realized a class of ``incommensurate DNA origami'' subunit, whose shape was superficially engineered to give rise to accumulating stretching of cohesive bonds upon 1D assembly \cite{berengut2020self}.  These studies indeed showed, through experiments and simulation, that frustrated subunits assembled into limited chains of fewer than $\sim 4-5$ subunits, in contrast to unfrustrated assemblies which exhibit unlimited 1D assembly lengths (i.e. exponentially distributed). While this proof of concept study demonstrated a range of conditions where frustration prevented unlimited aggregation, it remains to be understood for this, or any other experimentally designable assembly motif, what is maximum range of self-assembly sizes that can be reached, what are the range of thermodynamic conditions where these can be reached, and crucially, how are these controlled by particle-scale interactions, misfitting shape and deformability?

A critical challenge to answering this question is to predict the relevant mechanisms of {\it frustration escape} that mitigate the cost of misfit as the assembly grows~\cite{hagan_equilibrium_2020}.  Simply put, previously studied mechanisms of GFA exhibit a range of power-law accumulation of elastic stress at small-size, in which the elastic costs of intra-assembly strain grow superextensively with finite domain dimension~\cite{Meiri2021}.  Because the subunits and their interactions are ``soft'' at large domain sizes, large intra-assembly stress inevitably trigger distinct structural modes of (at least partial) frustration relaxation. For example, crystalline caps on spherical surfaces eventually reach a size where it is favorable to incorporate topological disclination defects that screen the far-field stresses of curvature and ultimately set an upper limit to the thermodyanmic costs of frustration~\cite{Li2019}.  Alternatively, even in the absence of defect formation, GFA models can undergo a mechanism of ``shape flattening'' where the assembly elastically deforms to a motif that is geometrically compatible with bulk assembly~\cite{Grason2016}.  In this case, the range of stress accumulation is delimited by the relative elastic costs to propagate intra-assembly stress vs. deforming the packing to a compatible one from its locally preferred misfitting one.

In this paper, we present and study a model of the planar assembly of ``warped-jigsaw'' (WJ) particles, shown schematically in figure \ref{fig:1}.  This misfitting particle was first proposed in \cite{grason_misfits_2017} where it was described heuristically as a simple means to illustrate how local mechanisms of shape frustration propagate to multi-particle dimensions in aggregates.  The two key ingredients of the particle design are the specific edge interactions that promote local orientational correlations (i.e. the top (right) edge of particles only binds favorably to the bottom (left) edge of adjacent particles) and the trapezoidal particle shape (characterized by taper angle $\phi \neq 0$) that promotes a favorable rotation of adjacent left-right edges. The tapered particle shape results in preferred curvature along horizontally assembled rows, $\kappa_0 \simeq 2 \phi/d$. As shown schematically for the tetrameric assembly in figure \ref{fig:1}(b), row curvature frustrates the 2D planar, crystalline packing of warped puzzle pieces. In this study, we develop a particle-scale description of cohesively binding WJ particles to study the mechanisms and modes of intra-assembly strain propagation as a function of misfitting particle shape, as well as the geometric design of inter-particle cohesive interactions. 

We map the particle-scale description onto a continuum elastic description for rectangular aggregates of variable size, and directly derive the continuum elastic parameters (moduli, preferred curvature) in terms of parameters that describe the discrete particle shape and interactions. We show that the multi-particle ground states of this WJ particle design (with isotropic binding on its edges) are anisotropic fiber morphologies with potentially finite multi-particle widths.  Using a combination of numerical simulation and continuum theory, we study the elastic modes of intra-assembly deformation as a function of assembly sizes and particle features.  Specifically, we trace the mechanism of shape flattening in widening ribbons to the interplay of between the elastic costs of unbending curved rows and the accumulating costs of shearing between adjacent rows of the assembly.  We show that the range of frustration accumulation can be controlled through the design of the inter-particle binding, specifically through a ``pull-push-pull'' arrangement of attractive and repulsive binding sites.  We find that the range of finite width equilibria can reach sizes that far exceed the particle width when the net repulsive strength approaches (but does not exceed) attractive contribution between bound edges.  This result shows how the effects of frustration on assembly depend not only on the degree of shape-misfit, but also crucially, on the deformability of assembly to distinct elastic modes.  Simply put, reaching large equilibrium self-limiting dimensions requires WJ designs that are simultaneously {\it stiff} to row (un)bending and {\it soft} to inter-particle shears.  

The remainder of this paper is organized as follows.  We first introduce the discrete model of WJ particles based on specifically interacting trapezoidal subunits, and show that its elastic ground states select ribbon-like morphologies which extend along the normal to the preferred bending direction.  We then analyze the continuum elasticity theory and describe the elastic equilibrium for infinite length ribbons.  We show that the characteristic ``flattening size'' is determined by the ratio of row bending modulus to inter-row shear modulus, and equivalently by the ratio of repulsion to attraction interaction strength in the binding design of WJ particles.  We then demonstrate how the repulsion/attraction interaction ratio at the particle-scale controls the range of self-limiting sizes that may be achieved, based on numerical energy-minimizations of prebuilt assemblies.  We conclude with a simple analysis of the thermal fluctuations of the finite ribbon widths of assemblies and a discussion of the ramifications of this finding on the design and implementation of GFA into synthetically engineered particle assemblies.

\section{Discrete WJ model and energetic ground states}

\label{sec: discrete}

Here, we describe the model of planar assembly of discrete WJ particles, numerical methods for computing planar ground state energies and the emergence of domain anisotropy from rectangular domain assemblies.  We connect this discrete-particle description to the continuum (multi-particle) level of aggregates in the subsequent section.

\subsection{Warped-jigsaw particle model}
The trapezoidal WJ  particles are modeled as rigid arrays of interaction sites of three different types:  $24$ volume exclusion sites ($5$ on each side, and $4$ on the corners), $8$ attractive sites ($2$ on each side), and $4$ repulsive sites ($1$ on each side). The attractive and repulsive sites have centers along a trapezoid, defining the nominal particle size $d$ (see~figure \ref{fig:1}). Volume exclusion sites are arrayed along the four edges of the inner trapezoid of height $d' = 0.645 d$, with spacing $0.161 d$. Exclusion sites interact with all other exclusion sites on other particles with a cut-off Lennard-Jones (i.e. WCA potential) interaction,
\begin{equation}\label{eq:1}
U_{\rm ex}(r)=u_{\rm ex}+u_{\rm ex}\Big[\Big(\frac{\sigma}{r}\Big)^{12}-2\Big(\frac{\sigma}{r}\Big)^6\Big] \ \ {\rm for} \ r\leq \sigma
\end{equation}
where $\sigma$ is the smooth cut-off of potential (i.e. $U_{\rm ex}(r> \sigma) = 0$), and $u_{\rm ex}$ parameterizes the energy scale of excluded volume interactions. For this study, $\sigma = 0.323 d$.

On each face, there are two attractive sites (shown in~figure \ref{fig:1}) offset from the midpoint of the face, and a single repulsive site at the side center,
%central pivot point in 
between the two attractive sites.  Attractive and repulsive site locations are pushed out from the inner array of exclusion sites by $\Delta = 0.177 d$ to avoid excluded volume contact for %ideally
bound particles under sufficiently small deformations. Interactions between attractive and repulsive sites are specific, meaning that interactions with a given site, are only non-zero between a particular site on the corresponding edge of an adjacent particle (shown schematically with complementary colors in Fig.~\ref{fig:1}) . For example, the upper attractive site of the particle's right edge only interacts with the upper attractive sites on the neighbor particle's left edges, and likewise, the central repulsive site on the top edge only interacts with repulsive sites on the bottom of edge of neighboring particles. Such specificity encodes a ``lock and key'' type binding of the type that can be readily engineered in DNA origami particles~\cite{Gerling2015, Sigl2021}, and for the purposes of the WJ particle maintains local orientational correlations between bound neighbors in the assembly (i.e. neighbor trapezoids can only bind in nearly `parallel' orientations, notwithstanding the rotations required by tapered shape).  Attractive and repulsive interactions are modeled by a soft, finite range potential. The attractive interactions are given by
\begin{equation} \label{eq:2}
     U_{\rm a}(r)=-u_0\Big(1+\cos{\pi\frac{r}{r_c}}\Big)  \ \  {\rm for} \ r \leq r_c ,
\end{equation}
where $r_c$ is the range on the specific interactions (beyond which $U_{\rm a}(r>r_c)=0)$ and $-2 u_0$ is the depth of the pair binding. Repulsive sites interact via an inverted form of the attractive potential, described by the potential
\begin{equation}\label{eq:3}
     U_{\rm r}(r)=-2RU_{\rm a}(r)  ,
\end{equation}
where $R$ is defined as the repulsive strength relative to the strength of the attractive sites. %Assuming
By keeping $R<1$, edge interactions remain net attractive and the lowest energy of a bound pair of particles is $-4u_0 (1-R)$. %ground state of bound edges is for associated attractive and repulsive sites at zero separation with an energy of $-4u_0 (1-R)$  

As shown in~figure \ref{fig:1}, $d$ is defined by the length of the top of the trapezoid defined by being the smallest trapezoid that encapsulates the excluded volume of the rigidly connected volume exclusion sites, which is the nominal particle size. In the limiting case $\phi=0$, particles arrange in a square lattice with spacing $d$.  The wedge angle $\phi$ quantifies the deviation from right internal angles, and thus parameterizes frustration in the WJ model.  Placing three ideally bound particles in horizontal rows as in figure \ref{fig:1}(c), leads to an effective curvature $\kappa_0$, that can be characterized by the radius of curvature of the particle centers,
\begin{equation}
\kappa_0^{-1} = \frac{d \sqrt{1 + \sin(2\phi)}}{2\tan{\phi}}
\end{equation}
Intuitively this curvature frustration vanishes linearly in the limit of small taper angles $\kappa_0 (\phi \ll 1) \simeq 2 \phi /d$.  

\begin{figure}[h]
    \centering
    \includegraphics[width=0.975\linewidth,height=\textheight,keepaspectratio]{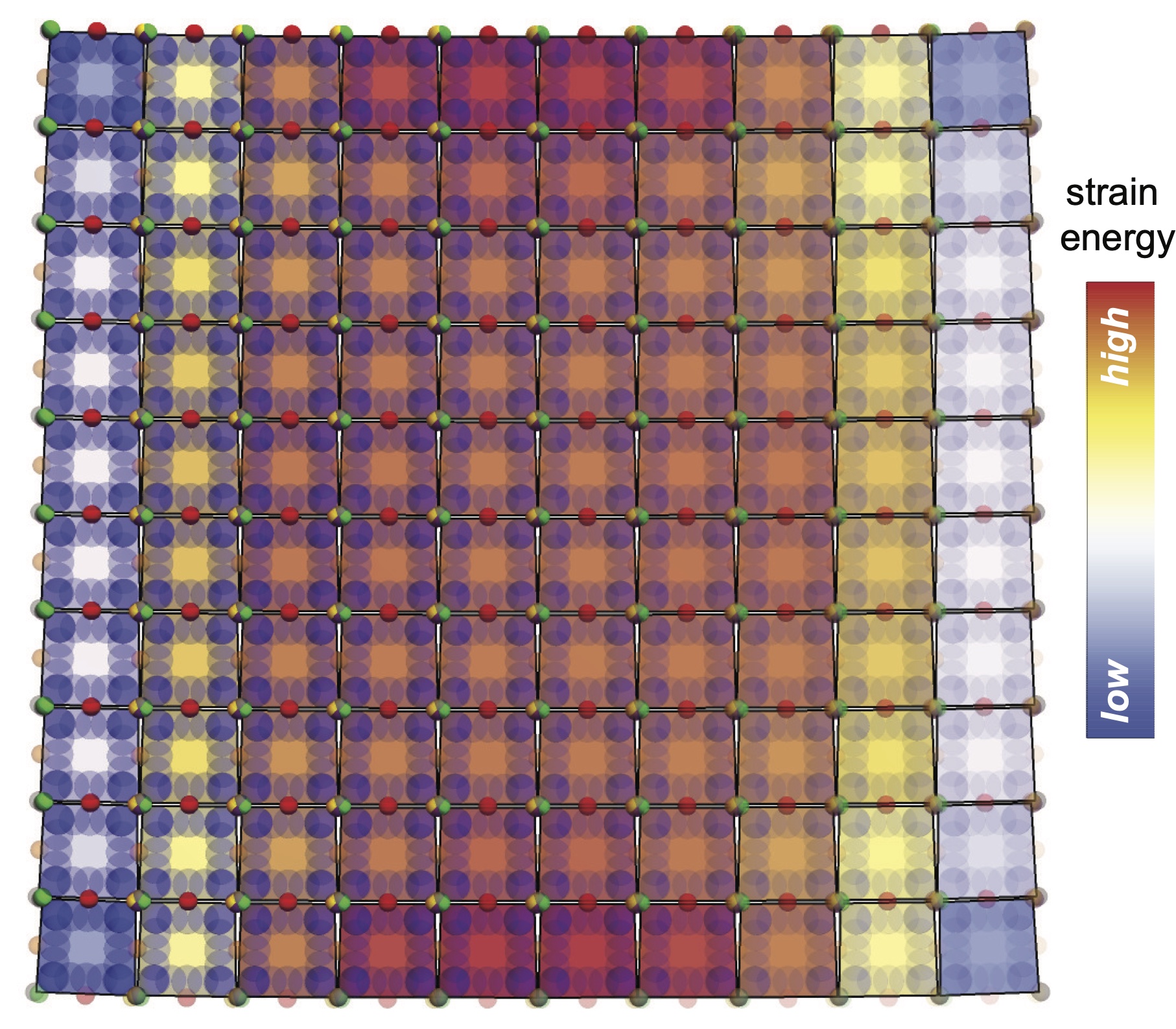}
    \caption{Frustrated assembly exhibits gradients in stress, evidenced by the excess energy density in a numerically relaxed $10 \times 10$ assembly, for the case $\phi = 0.03$, $t/d = 0.968, r_c/d = 0.323$, and zero repulsive interaction strength $R=0$. The interaction-site representation is overlaid with a representation of particles as trapezoids, colored by their respective interaction energy in excess of an unfrustrated reference state.}
    \label{fig:assembly-example}
\end{figure}

Unless otherwise stated, the geometric parameters used in this study are wedge angle $\phi = 0.03$, and elastic thickness parameter $t/d = 0.968$, with varying interaction parameters $r_c/d, R$.  While vertical and horizontal 1D rows of bound particles are unfrustrated (figure \ref{fig:1}(c) and (d)), assembly into larger 2D clusters requires strains in the assembly, as illustrated by the variable energy density in the simulated square cluster in~figure \ref{fig:assembly-example}, obtained via numerical minimization described next.

\subsection{Numerical optimization of multi-particle domains}

To predict the propagation of particle-scale misfit and interaction to patterns of intra-aggregate deformation and their associated energetics, we construct and study numerically minimized arrays of WJ particles.  In short, we minimize the %zero temperature free energy 
total interaction energy of rectangular arrays (of dimensions $W \times H$) and periodic arrangements of particles using a conjugate gradient algorithm (implemented  with the SciPy function optimize.minimize~\cite{2020SciPy-NMeth}), beginning from an initial (pre-stressed) regularly spaced 2D rectangular lattice of aligned particles arranged in cohesive contact. 
%The gradient of the free energy of the array of particles as a function of $x_i$, $y_i$, and $\theta_i$, where $i$ runs from $1$ to the total number of particles, was computed analytically and the gradient of the free energy of the periodic arrangement of particles was computed numerically using finite difference. The step size used in the finite difference algorithm was the square root of machine error, $\sqrt{\epsilon_{machine}}\approx1.5\times10^{-8}$. 
The numerical minimization terminated when the norm of the gradient of the free energy was less than $10^{-5} u_{\rm ex}/d$, while the %free 
total energies involved were never less than $u_{\rm ex} \times10^{-1}$. 

In the initial (uncurved) reference state, bonds between horizontal neighbors are pre-strained, wedged open by an amount that depends on the taper angle $\phi$. Given this pre-strained state, for sufficient short attractive ranges, energy barriers can prevent the conjugate gradient algorithm from reaching a stable aggregate states which maintain cohesive throughout the assembly.  To account for this effect of finite interaction range, we perform minimization in two ways dependent on the range of attractions, with method used for each set of data is summarized in Appendix \ref{sec: parameters}.

\noindent \textbf{Method 1:} When $r_c$ is sufficiently large compared to the inter-attractive site gap in the uncurved 2D lattice, we relax via the conjugate gradient algorithm from this 2D reference state: a rectangular array of particles defined by a center to center difference in the $x$-direction, $\Delta x=d(1+2\tan\phi)$, a center to center difference in the $y$-direction, $\Delta y=d$, and with all particles aligned to the lattice directions. For the periodic arrangement of particles (for modeling infinitely tall, $H \to \infty$, ribbons), the initial state was defined in the same way, based on a periodic vertical stacking of a single row of the rectangular lattice with optimized vertical spacing. 

\noindent \textbf{Method 2:} When the ultimate target value of $r_c$  is small compared to the inter-attractive site gap in the uncurved reference state, we first apply Method 1 for the same particle array but with the binding range artificially larger than the target value and sufficiently large to provide cohesion between neighbors in the uncurved reference state.  The output of the larger-$r_c$ equilibration is used as the starting point for the next minimization, and the process is iterated until the binding range is ramped down to its target value. 

In all final configurations, we analyze the energetics by normalizing the energy per particle relative to a bulk state with uncurved rows, denoted as $\epsilon_\infty$, a value that can be computed to an excellent approximation using the ``flattening energy" described in the continuum theory in the following section.  We also define the local elastic energy of particle $i$ as $1/2$ of the bond energy of $i$ with its $z_i$ neighbors minus the energy of $z_i$ perfect cohesive bonds (i.e. $-4 z_i R u_0$).  This results in spatial maps of the elastic energy density as shown in figure \ref{fig:assembly-example}, where particles are most strained in the center of the particle and tend to relax toward the free boundaries of the aggregate, particularly its corners, due to the fewer geometric conflicts with neighbors.  Following the description of GFA introduced in ref. \cite{hagan_equilibrium_2020} we refer to this elastic energy per particle as the {\it excess energy}, $\epsilon_{\rm ex}$, as it is the energetic cost of assembly in excess of the per particle energetics of ideal (strain-free) cohesion.

\subsection{Energy landscape and ribbon morphology selection}

\begin{figure*}
    \centering
    \includegraphics[width=0.85 \linewidth,height=\textheight,keepaspectratio]{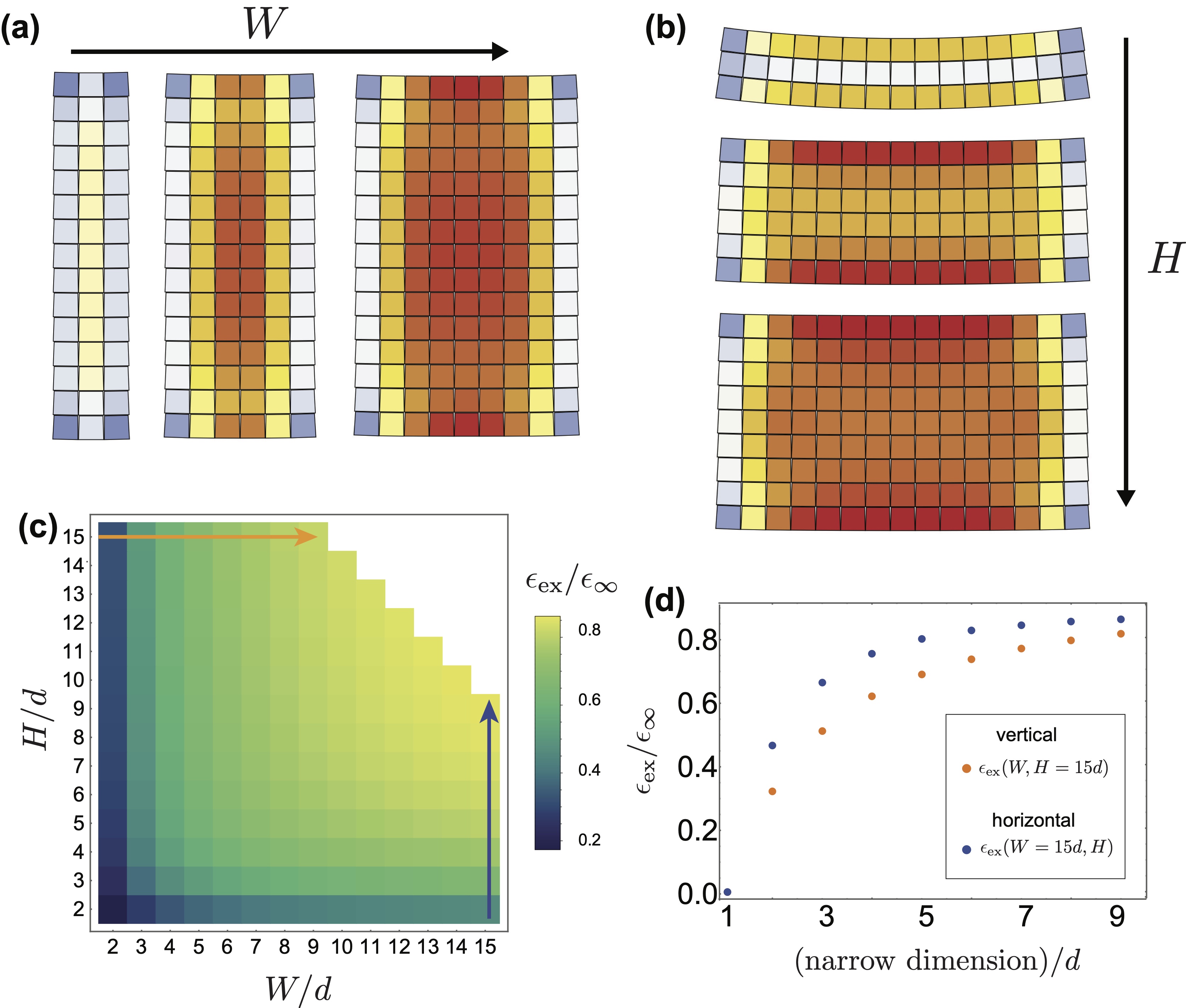}
    \caption{(a) Vertical ribbons, numerically minimized structures with constant length of $L/d = 15$ and increasing width $W$ are shown. (b) Horizontal ribbons of fixed $W/d = 15$ and with dimension $L$ increasing. (c) Mean excess energy density landscape as a function of assembly width and assembly length for $r_c/d = 0.323, R = 0$ for rectangular WJ assemblies that are less that 150 total units (larger sizes correspond to white region). (d) A comparison of excess energy vertical vs. horizontal structres for equal total size as a function of the narrow dimension.  In these plots, the large dimension is fixed to 15 units in size as indicated by the orange and purple arrows in (c).
    %Here, we can see the anisotropy of the mean elastic energy density, which comes from the anisotropy of the trapezoidal particles. 
    %We also note that vertical ribbons simulated with periodic boundary conditions have lower energy density than their finite counterparts, which we show in~\sref{inf ribbon vs finite ribbon}, implying that this trend is generic. 
    %I moved this last bit to the text
    }
    \label{fig:vertical_vs_horizontal}
\end{figure*}

Based on the WJ model introduced above, we may in general consider the energetics (i.e. the $T=0$ thermodynamics) of aggregated states of different size and shape.  In this study we focus on the branch of aggregate structures with the lowest energy density (when edge interactions are equal on all sides), which take the form of finite-width, "vertical" ribbons.  The focus on this morphology can be understood by analyzing the landscape of excess energy in Fig. figure \ref{fig:vertical_vs_horizontal}, which is computed for aggregates of variable width $W$ and height $H$, for WJ particles with $r_c/d = 0.323, R = 0$.
%\GMG{We need to list the parameter values use for these calculations}.
This landscape shows that the elastic cost increases with either increased $W$ or $H$, due to the accumulation of inter-particle elastic stresses.  Notably, the excess energy landscape is not symmetric with respect to $W$ and $H$, as the preferred direction of curvature imposes a strong anisotropy in the nature and magnitude of intra-aggregate stress that propagate in the distinct directions. ``Tall and narrow'' aggregates ($H \gg W$) form straight vertical ribbons of stacked curving rows, with elastic energy predominately varying in the width directions, with the exception of narrow boundary regions at the top and bottom free ends of the ribbon. Likewise, ``short and wide'' aggregates ($H \ll W$) from arched horizontal ribbons, with a mean curvature that tends to flatten with increased height and with energy density that varies predominantly in the height direction.

The total energy density for this assembly, can be written as
\begin{equation}
\label{eq: epsilon}
    \epsilon(W,H) =\epsilon_0+  2 \Sigma \Big( \frac{1}{W}+ \frac{1}{H} \Big) +  \epsilon_{\rm ex}(W,H) ,
\end{equation}
where $\epsilon_0$ is the (strain free) per particle cohesive gain of particles in the bulk and $\Sigma$ is a edge energy penalty accounting for the deficit of cohesive bonds at the free ends of the rectangular cluster.  Notably the thermodynamic tendencies of the edge energy favor  large (unlimited size) aggregates to minimize the fraction of particles at the boundary, a tendency which competes with the elastic costs of frustrations.  

Analysis of these total energy landscapes reveals for low enough $\Sigma$ the minima of the total energy landscape fall into two branches, one of which corresponds to channel of finite width vertical ribbons, while the second corresponds to the to a channel of finite height horizontal ribbons.  The absolute local minimal of these branches correspond to the case of infinite height ribbons in the first case and the end-free or fully closed annular ribbons in the second case.  In both cases, the favored structure is effectively symmetric along its long direction and has open boundaries only perpendicular to the long axis. The tendency for frustration to lead to highly-anisotropic morphologies is apparently generic~\cite{Grason2016}, driven by the fact that frustration introduces gradients in the packing but those gradients need not extend in all directions in the assembly.  By keeping one of the assembly dimensions finite, the ground state can at least partially ``escape frustration'' by extending a finite width motif to an infinite extent.  For example, this mechanism has been observed and studied in models of crystallization~\cite{Schneider2005, Meng2014} on spherical surfaces, twist-frustrated filament bundles~\cite{Hall2016} and flexible polygon assemblies~\cite{Lenz2017, Meiri2021}.  

In the present case of WJ particles, the symmetry of the particle introduces an effective polarity to the assembly that breaks the symmetry between the different ribbon morphologies.  In figure \ref{fig:vertical_vs_horizontal} we compare the excess energy density vertical (straight) vs. horizontal (arched) ribbons as a function of the finite dimension for the same fixed large dimension, i.e. height and width respectively.  These show that the excess energies of both states vanish for narrow dimensions and asymptotically approach the same bulk cost $\epsilon_\infty$ for large dimensions, and that the elastic cost of frustration of vertical ribbons is always less than the horizontal ribbons.  As the horizontal and vertical edges have the same surface energy in this model, this implies that if the thermodynamic ground state of the assembly is finite in any dimension, it will be a vertical ribbon.  Below, we show that the nature of the elastic energy accumulation with increasing width of vertical ribbons derives from the costs of inter-layer shears, while it can be intuitively understood that horizontal ribbons are essentially ``bent 2D crystals'' and must also generate elastic costs of differential compression and dilation of the horizontal rows.  Evidently, as we will describe in detail elsewhere, the distinct kinematics of frustration propagation normal to vs. along the curvature direction accounts for distinct energetics of stress propagation in those directions of assembly, biasing assembly thermodynamics to the minimal-energy vertical ribbon morphology.  For the remainder of this article we focus our analysis on the mechanics and thermodynamics of finite-width vertical ribbon assembly in the WJ model.

%We claim that that the favored morphology of an assembly of our particles is to filament in the vertical direction to create long ribbons. We also claim that these vertical ribbons tend to grow in their long direction indefinitely. The anisotropy of the particle geometry leads to vertical ribbons being the preferred growth direction, as can be seen in~figure \ref{fig:8}. Additionally, the ribbons simulated with periodic boundary conditions have lower energy than finite ribbons with the same parameters, which is shown in~\sref{inf ribbon vs finite ribbon}.

\section{From discrete particles to continuum elasticity}

\label{sec: continuum}

Underlying the complex, multi-scale thermodynamics of GFA is the propagation of inter-assembly gradients and stress, whose elastic costs compete with cohesion to limit assembly size.  In this section, we describe the continuum theory of planar arrays of WJ particles, tracing the parameters that govern the stress to the particle shape and interactions.  We then describe the equations of mechanical equilibrium and their exact solution in the limit of finite width (vertical) ribbon domains, and deduce the mechanics of the {\it shape-flattening} transition between accumulating and saturated elastic costs with domain width.  

\subsection{Pairwise particle elasticity}\label{micro sec}

\begin{figure}[h]
    \centering
    \includegraphics[width=\linewidth,height=\textheight,keepaspectratio]{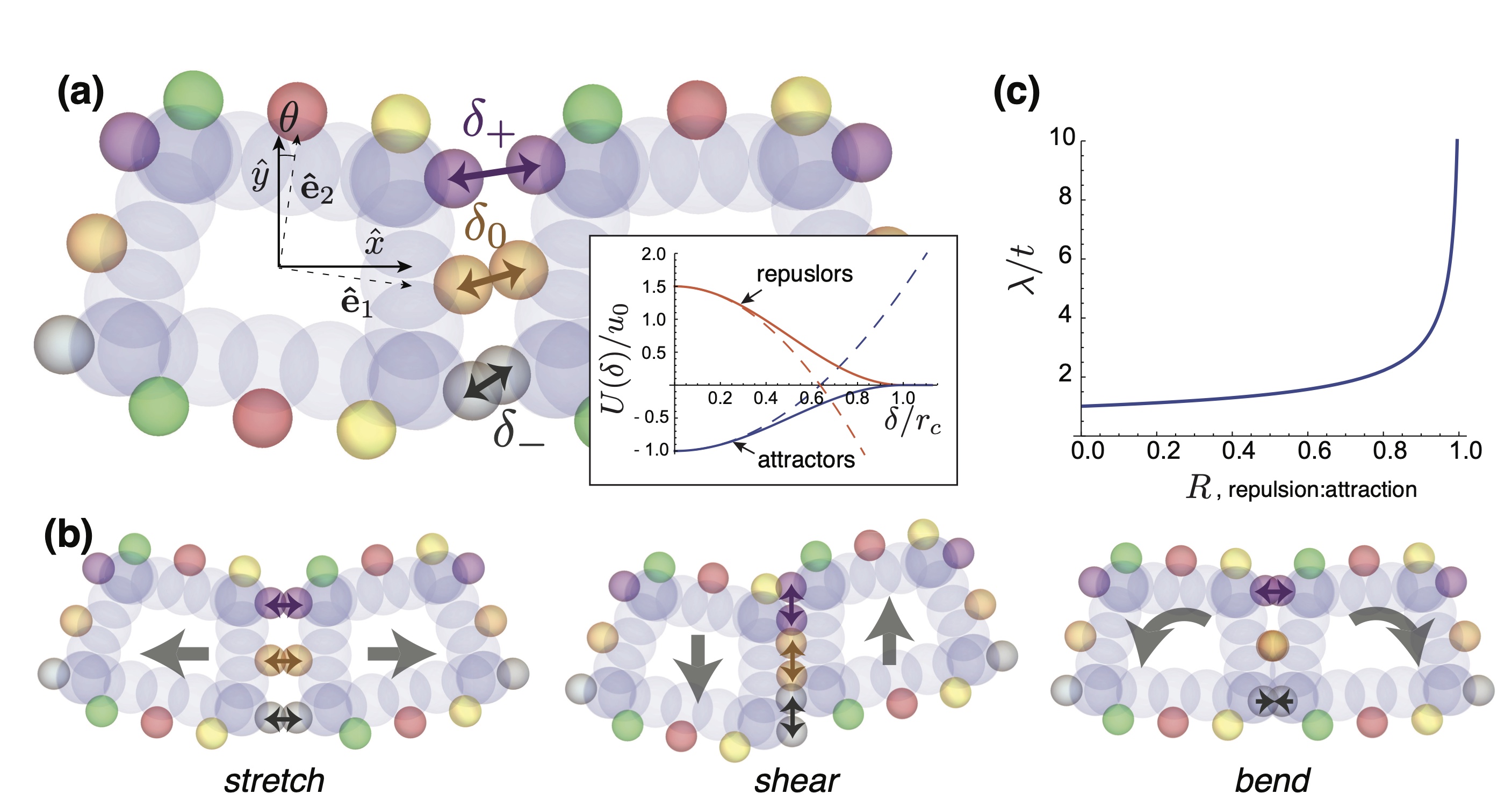}
    \caption{(a) Schematic of particle coordinate frame, respective attractive site distances $\delta_{+},\delta_{-}$ and repulsive site distance $\delta_{0}$. Inset graph shows the interaction potentials along with their second order expansions. (b) Schematic of stretch, shear and bend deformations and the resultant separations of interaction sites. (c) Plot of the characteristic flattening length scale $\lambda/t$ normalized by attractive site separation $t$, as a function of the relative repulsive/attractive interaction strength $R$. }
    %(b) The eigenvalues of the normal modes of a two particle system, equal to the continuum bending and shear moduli, are plotted as a function of relative repulsive strength $R$. Inset shows the resultant dependence of the length scale $\lambda$ that diverges with $R$. (c) The two particle ground state has both attractive sites and the repulsive site completely overlapping. (d) The bending mode of deformation results in compressive and stretching separations of the attractive sites while the repulsive sites remain overlapping. (e) Shear and stretching modes of deformation have all three interaction sites undergoing equal displacements. }
    %(b,d-f) The two particle ground state has both attractive sites and the repulsive site completely overlapping. The normal modes of the two particle elastic energy are stretching, bending, and shear. (c) The eigenvalues of the normal modes are plotted along with the continuum elastic moduli as a function of repulsive strength relative to cohesive strength.}
    \label{fig:pair}
\end{figure}

We begin with a discrete rectangular lattice of particles: we assign each particle horizontal and vertical indices $(m,n)$, corresponding particle center coordinates are denoted $\rv_{m,n} \equiv (x_{m,n}, y_{m,n})$, and particle orientations are determined by angle $\theta_{m,n}$ with reference to a global frame defined by the lattice directions, as shown in~figure \ref{fig:pair}(a). The orientation of the rigid particle is described by a local orthonormal frame $\{\ev^{m,n}_1, \ev^{m,n}_2 \}$, where
\begin{multline}
    \ev^{m,n}_1 =\cos(\theta_{m,n}) \hat{x} - \sin(\theta_{m,n}) \hat{y};  \ev^{m,n}_2  \\ = \sin(\theta_{m,n}) \hat{x} + \cos(\theta_{m,n}) \hat{y};
\end{multline}
describe the local base and height directions of the $(m,n)$ trapezoidal particle. We consider sufficiently small local strains from the reference configuration above such that we may approximate inter-particle bond energies by harmonic distortions from the local minima (maxima) of attractive (repulsive) interactions.  The existence of such a small strain limit is achieved in the $\phi \to 0$ limit of vanishing frustration.  Given the form of the interaction potentials, equations \ref{eq:2} and \ref{eq:3}, this takes the form,
\begin{equation}
    U_{\rm edge} \simeq -2 u_0 (1-R) + \frac{u_0 \pi^2}{2 r_c^2} \big(\delta_{+}^2+\delta_{-}^2-2R\delta_{0}^2 \big)
\end{equation}
where $\delta_{\pm}$ refer to Euclidean separation between the two attractive sites and $\delta_0$ is the corresponding distance between the central repulsive site, as shown in ~figure \ref{fig:pair}(a). In the elastic description, we replace these soft interactions by harmonic springs with stiffness $k_{\rm a}=u_0\pi^2/r_c^2$ and $k_{\rm r}=-2Ru_0\pi^2/r_c^2$, for attractive and repulsive sites respectively (inset of  ~\ref{fig:pair}a).
%The interactions between the particles are modeled with nearest neighbor interactions on the lattice and define the linear elastic energy as the expansion of the interactions to second order, where the spring constant for the attractive sites is $k_a=u_0\pi^2/r_c^2$, and the spring constant for the repulsive sites is $k_r=-2Ru_0\pi^2/r_c^2$. We define a local frame for each particle represented in the basis of the global frame as,
%\numparts
%\begin{equation}\label{eq:4a}
%\hat{e}_1^{m,n}=\mathbf{R}_{\theta_{m,n}}\hat{x}=\big %[\cos(\theta_{m,n}),\sin(\theta_{m,n}))\big]
%\end{equation}
%\begin{equation}\label{eq:4b}
%\hat{e}_2^{m,n}=\mathbf{R}_{\theta_{m,n}}\hat{y}=\big[-\sin(\theta_{m,n}),\cos(\theta_{m,n})\big]
%\end{equation}
%\endnumparts
Based on the harmonic approximation of bound neighbors, the elastic energy of horizontal and vertical neighbors (which correspond, respectively, to neighbors in $\hat{x}$ and $\hat{y}$ direction of reference 2D lattice) take the form
\begin{multline}\label{eq:5}
E_x(m,n) = \frac{(2 k_{\rm a} - k_{\rm r})}{2}\big|\Delta_x \rv -d\langle\ev_1\rangle_x\big|^2 \\ +\frac{k_{\rm a} t^2}{4}\big|\Delta_x\ev_2+2\phi\langle\ev_1\rangle_x\big|^2 + {\cal O}(\phi^4)
\end{multline}
and 
\begin{equation}\label{eq:6}
    E_y(m,n) = \frac{(2 k_{\rm a} - k_{\rm r})}{2}\big|\Delta_y \rv -d\langle\ev_2\rangle_y\big|^2+\frac{k_{\rm a} t^2}{4}\big|\Delta_y\ev_2\big|^2 ,
\end{equation}
where $\Delta_i$ refer to the discrete difference between sites along $i=x,y$ directions and $\langle \cdot \rangle_i$ refer to the mean value taken over the neighbor pair.  The first terms in $E_i$ favors uniform center-to-center spacing along the (mean) edge orientation equal to the particle size, whereas the second terms represent angular interactions.  Binding over vertical faces favors parallel orientations, while the tapered edge shape introduces a preferred rotation by $2 \phi$ of horizontal neighbors.
%Where $\vec{u}$ is the displacement, $\vec{r}$ defines the lattice coordinates, $Y=2u_0\pi^2(1-R)/r_c^2$, and $B=t^2u_0\pi^2/2r_c^2$. $\langle\rangle_x$ represents a mean, and $\Delta_x$ represents a difference between particles $m+1,n$ and $m,n$. The elastic energy for two neighbors in the $y$-direction is,
%\begin{equation}\label{eq:6}
%    E_y=\frac{1}{2}\Big(Y||\Delta_y\vec{r}+\Delta_y\vec{u}-t\langle\hat{e}_2\rangle_y||^2+B||\Delta_y\hat{e}_1||^2\Big)
%\end{equation}
%Where $\langle\rangle_y$ represents a mean, and $\Delta_y$ represents a difference between particles $m,n+1$ and $m,n$. 
The full elastic energy is then given by summing over the 2D reference lattice of the particles,
\begin{equation}\label{eq:7}
    {\cal E}_{\rm elas}=\sum_{m,n}E_x(m,n)+E_y(m,n).
\end{equation}
For the present study, we consider domain shapes of rectangular shape (i.e. $M$ columns $\times N$ rows).

\subsection{Continuum elasticity of WJ domains}
We transform our discrete particle description to a continuum limit, based on the assumption that domains are sufficiently large to be described by smooth functions of the reference coordinates $\xv_{m,n} \equiv d (m  \hat{x} + n  \hat{y}$).  We take this limit for a function $f$ of the reference coordinates  as $f_{m \pm 1,n \pm 1}\to f(x \pm d,y \pm d)$ and $\Delta_i f \to d \partial_i f$  followed by taking the limit that particle dimensions are infinitesimal compared to the domain size. We define the elastic energy as a function of particle orientation $\theta(\xv)$ and particle displacement,
\begin{equation}
    \uv(\xv) = \rv(\xv) - \xv ,
\end{equation}
where $\xv$ is the particle position in the reference state and $\rv(\xv)$ is its position in the deformed state.  The continuum elastic energy takes the compact form,
\begin{multline}\label{eq:continuum}
    {\cal E}_{\rm elas} \big[\uv(\xv), \theta(\xv) \big] =\frac{1}{2}\int_\Omega d^2 \xv \Big\{Y\big(\partial_i  u_j - \epsilon_{ij} \theta \big)^2 \\ +B\big(\nabla_\perp\theta-\kappa_0\hat{x}\big)^2\Big\}
\end{multline}
where $\epsilon_{ij}$ is the 2D Levi-Cevita tensor (we follow standard Einstein summation conventions of indices), $\nabla_\perp = \hat{x} ~ \partial_x + \hat{y} ~ \partial_y $ is the planar gradient and $\Omega$ is a rectangle domain on the reference lattice $\Omega=[0,W]\times[0,H]$.  Notably the coarse-graining introduces three continuum parameters:  the preferred row curvature $\kappa_0=2\phi/d$, the {\it stretch/shear modulus},
\begin{equation}
\label{eq:Y}
    Y \equiv  2k_{\rm a}- k_{\rm r} = \frac{2 u_0 \pi^2}{r_c^2}(1-R)
\end{equation} and the {\it bending modulus} of particle rows
\begin{equation}
\label{eq:B}
    B \equiv  \frac{k_{\rm a} t^2}{2} = \frac{u_0 \pi^2 t^2}{2 r_c^2} .
\end{equation} 
We consider the implications of the dependence of these effective moduli on particle interaction parameters for the equilibrium intra-particle stress distributions below.

%The mechanism of frustration is clear from a simple inspection of continuum elasticity energy in eq. (\ref{eq:continuum}).  
The first term in eq. (\ref{eq:continuum}) penalizes deformations that disrupt the local 2D crystalline packing, including intra-row stretching/compression and inter-row shears.  The coupling of positional displacements to particle orientations is precisely of the form that preserves rotational invariance of the elastic energy.  A rotation by constant $\delta \theta$ is described by $\theta(\xv) \to \theta(\xv) + \delta \theta$ and $u_i (\xv) \to u_i + \delta \epsilon_{ij} x_j$ (for small enough rotations), such that the local strain $\partial_i u_j - \epsilon_{ij} \theta $ is invariant to ${\cal O}(\delta \theta^2)$.  Hence, this term favors particles co-oriented to lattice directions. The second term favors {\it uniform gradients} of particle orientation, $\nabla_\perp \theta = \kappa_0 \hat{x}$, which implies a favored local configuration with straight vertical columns and constant curvature along horizontal rows.   While uniform, global rotations introduce no lattice strain, variable local rotations favored by gradient of $\theta(\xv)$ are effectively {\it Goldstone modes} of the frustration-free elasticity~\cite{Chaikin95}, and hence require an elastic cost which can accumulate with domain size. 

Variational analysis of the elastic energy functional %, eq. \ref{mech eq}), 
gives the following conditions for mechanical equilibrium in the interior of the domain: force balance gives
\begin{equation}
    \nabla_\perp^2 u_i = \epsilon_{ij} \partial_j \theta ,
\end{equation}
and torque balance gives
\begin{equation}
   B \nabla_\perp^2 \theta = Y \big(\nabla_\perp \times \uv + 2 \theta \big) ,
\end{equation}
where $\nabla_\perp^2$ and $\nabla_\perp \times$ are the respective Laplacian and (2D) curl operators on the plane.  Additionally, free boundary conditions require %the following conditions at boundary to the domain 
that, at the domain boundary denoted $\partial \Omega$ with local normal direction ${\bf n}$: vanishing normal stress gives
\begin{equation}
\label{eq: stressBC}
    n_i \big( \partial_i u_j - \epsilon_{ij} \theta\big)\Big|_{\partial \Omega} = 0 ,
\end{equation}
and vanishing torque gives
\begin{equation}
\label{eq: torqueBC}
    \big[ ({\bf n} \cdot \nabla_\perp) \theta -  \kappa_0 n_x \big]\Big|_{\partial \Omega} = 0 .
\end{equation}

%As described in Sec. \ref{sec: discrete} above, our primary interest in this study is the thermodynamic ground states, which correspond to vertically oriented and infinitely long ribbons ($H \to \infty$), which are  potential finite (i.e. self-limiting) widths.  
Based on the results in section \ref{sec: discrete} above, we focus on the ground state energetics of the bulk state and of finite-width vertical ribbon morphologies ($H \to \infty$ in both cases). As shown for the finite ribbon simulations in Fig. \ref{fig:vertical_vs_horizontal}, %mechanical equilibrium becomes 
equilibrium stresses are uniform along the length of the ribbon (i.e. sufficiently far from its free ends) and hence, we consider the solutions which are uniform in $y$.  We solve the equations of mechanical equilibrium for a ribbon of width $W$, %in~\ref{infinite ribbon}, 
which yield the following profiles for displacements and particles orientations for $x\in[0,W]$:
\begin{equation}
    \theta_{\rm eq}(x) = \kappa_0 \lambda \Big[ \frac{ 1- \cosh (W/\lambda) }{\sinh (W/\lambda)} \cosh (x/\lambda) +\sinh (x/\lambda) \Big] ,
\end{equation}
\begin{equation}
    u^{\rm eq}_y(x) =  -\kappa_0 \Big[ \frac{ 1- \cosh (W/\lambda) }{\sinh (W/\lambda)} \sinh (x/\lambda) +\cosh (x/\lambda) -1\Big] ,
\end{equation}
and $u^{\rm eq}_x =0$, where the characteristic length scale $\lambda$ is defined by the ratio of bending to stretch/shear moduli
\begin{equation}
    \lambda \equiv 2 \sqrt{\frac{B}{Y} } .
\end{equation}
These displacements can be evaluated in ${\cal E}_{\rm elas}$ to map the local elastic energy distribution (i.e. local energy density at $\xv$)
\begin{equation}
\label{eq:felas}
     f_{\rm elas} (\xv) \equiv \frac{Y}{2} \theta_{\rm eq}^2(x) + \frac{B}{2} (\partial_x \theta_{\rm eq}(x)-\kappa_0)^2
\end{equation}
  which we analyze below, and further, to compute the per particle excess (elastic) energy due to frustration,
\begin{equation}\label{eq:excess}
    \epsilon_{\rm ex}(W)=\frac{1}{2}B\kappa_0^2\Big(1-\frac{\lambda}{W}\tanh{W/\lambda}\Big) .
\end{equation}
In the following section, we compare this continuum prediction of internal stress in WJ-ribbons to discrete particle minima and discuss the mechanical origins and implications of its complex width dependence.

\section{Intra-ribbon stress, accumulation and shape-flattening}

\label{sec: results}

\begin{figure*}
    \centering
    \includegraphics[width=0.95 \linewidth,height=\textheight,keepaspectratio]{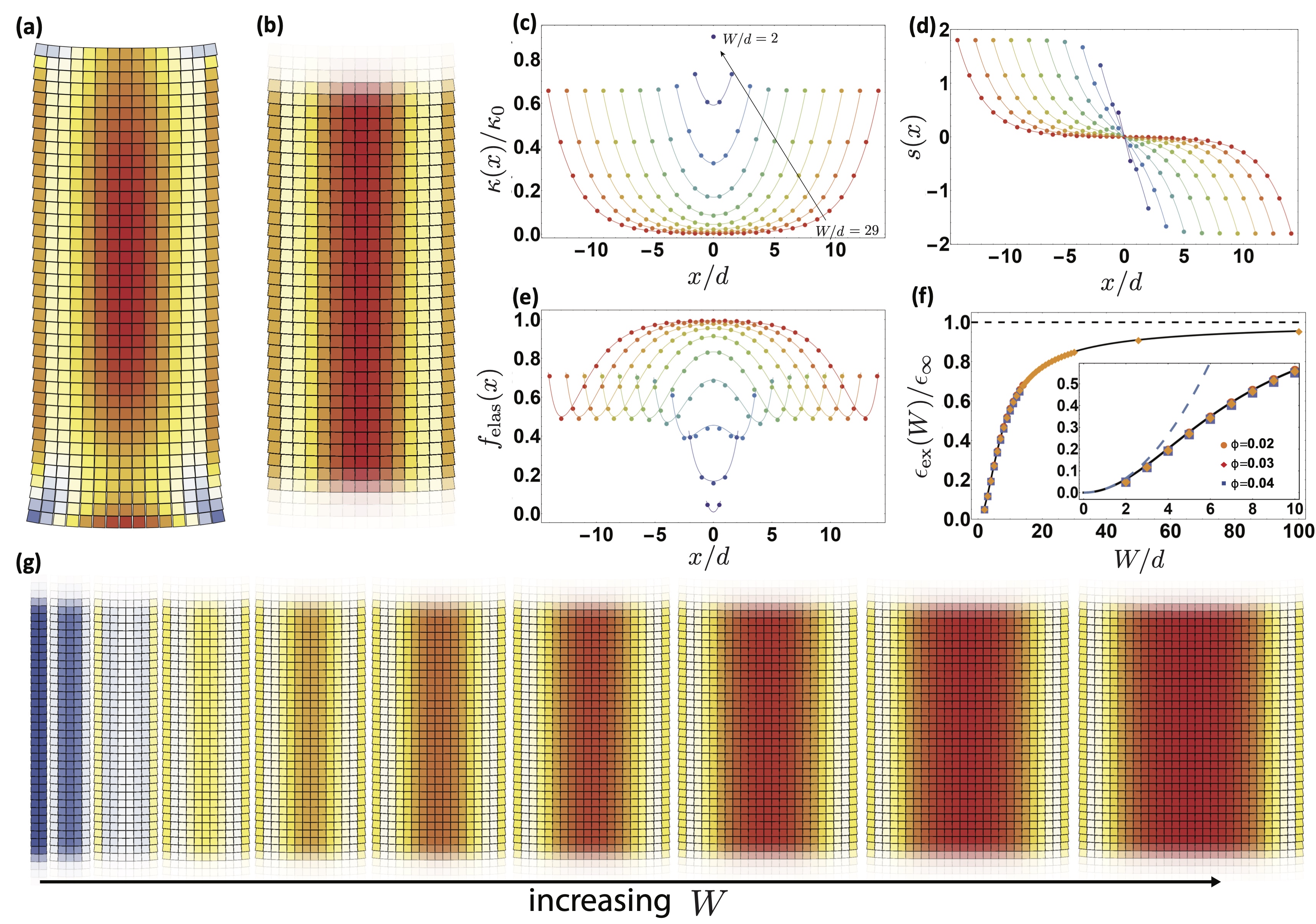}
    \caption{(a) A numerically minimized, finite length ribbon with $L/d = 40$ and $W/d = 15$ with strain energy density profile indicated by colors of the particles. %approximates the continuum model well except for the end-caps. 
    (b) An infinite-length and finite width $W/d=15$ ribbon simulated with periodic boundary conditions. The strain energy density distribution is close to that of in the finite ribbon in (a), sufficiently far from the finite ribbons top and bottom boundaries.%, we visualize the ribbon with a length of $40$ sub-units to visually compare to (a). 
    (c) The row curvature distribution of an infinite ribbon normalized by the preferred row curvature, where $x$ is the particle position in the assembly. The curves are from continuum theory and data points from infinite-ribbon numerics. The sequence of colors are for different assembly width $W/d$, showing that particles near the center flatten for wider assemblies.
    %This highlights how the ribbon is geometrically frustrated. Notice that as the ribbon becomes wider, the center of the ribbon flattens, we call this the flattening transition. The data points are from the numerical energy minimization, and the solid curves are from the generalized elasticity theory. 
    (d) The particle shear distribution for a series of assemblies with varying width. %with its $y$-neighbors shows that there is a consistent curved boundary layer for any ribbon width. \DMH{some of these commented-out bits could be in the body text, but I don't think they can properly fit in the captions if they aren't clear from looking at the figures}
    (e) The elastic energy density contains contributions from the strains in (c) and (d), normalized by the value at which the energy density saturates for particles near the center of wide ribbons.
    %, leading to the growth of a constant energy density on the interior with a lower energy density boundary layer. The flattening transition can be seen here as the point when the interior of the ribbon reaches a constant energy density. 
    (f) The mean elastic energy density captures the flattening transition in its approach to an asymptote defined by the energy cost of the flattening of a row of a ribbon. The inset shows the small size power-law growth of the mean elastic energy density, before the flattening transition. These simulations were run with $R=0.95$ and $r_c/d=0.417$ from $W/d=2\to30$. 
    %$u_0=1.85$
    %\DMH{What parameters for all of the data and structures here? } 
    (g) Visualization of the elastic energy distributions in ribbons of increasing width.}
    \label{fig:width}
\end{figure*}

We combine the continuum elasticity decsription developed in the previous section with discrete particle numerics to analyze the mechanics of inter-particle stresses that emerge from local misfit.  %As described above, we focus our attention on infinitely tall, finite-width ribbons.  
%In particular, we consider only the patterns of deformation in the interior of the ribbon.  
As shown in ~figure \ref{fig:width}(a), for finite-length vertical ribbons, sufficiently far away from the upper and lower boundaries,  deformations are nearly uniform along the length and well modeled by the periodic (i.e. end-free) equilibrium, shown in~figure \ref{fig:width}(b). 

We first consider a case for fixed interaction parameters, specifically $R=0.95$, and analyze the variation of internal packing and elastic energy for ribbons of variable finite width.  Results in figure \ref{fig:width}(c-e) show the elastic distortion of ribbons varying for widths of 2 to 30 particles for a fixed taper of $\phi = 0.03(\simeq \kappa_0 d /2)$.  The corresponding discrete-particle ribbons are shown in figure \ref{fig:width}(g), with spatial maps of equilibrium elastic energy, $f_{\rm elas}(\xv)$, highlighted in heatmap, with blue and red indicating, respectively, low and high magnitudes of elastic strain.  The sequence of increasing ribbon width illustrates the effect of accumulated elastic costs of frustration, with magnitude of $f_{\rm elas}$ increasing with width.  For large ribbons, the elastic energy concentrates and saturates in magnitude in the inner region of the ribbons flanked by partially relaxed zones near the free boundaries.

The spatial organization of the mechanical equilibria and their variation with ribbon width is analyzed in figure \ref{fig:width}(c) and (d), which show, respectively the row curvature $\kappa_{\rm eq}(x)$ and the xy shear strain $s(x) \equiv (\partial_y u_x + \theta) = \theta_{\rm eq} (x)$ for ribbons of increasing width.  We note that (fit-free) comparison between the continuum theory and corresponding values of the discrete WJ particle ground states show remarkable quantitative agreement, even for ribbons of only very few particles in width.  For relatively narrow ribbons, equilibria tend to maintain significant row curvature along their lengths, $\kappa_{\rm eq}(x) \approx \kappa_0$, which in turn leads to a nearly linear shear profile throughout the width.  For sufficiently wide ribbons, the equilibrium structure adopts a two-zone structure.  In the interior, rows are unbent (i.e. ``flattened'') and correspondingly shear strains are nearly eliminated from this ``defrustrated'' zone.  Near the free edges of large ribbons, rows bend up approaching their preferred curvature as required for vanishing torque at the free boundary, eq. \ref{eq: torqueBC}, which in turn leads to a build up of shear strains in these boundary layers.

It is straightforward to rationalize this behavior with a simple analysis of the 1D elastic energy density in eq.(\ref{eq:felas}) based on the value of row curvature at the center of the ribbon, $\kappa_c$.  Ignoring, for simplicity, the variation of curvature along the width, row curvature leads to a linear shear profile $s(x) \simeq \kappa_c x$ and corresponding shear elastic energy density $\sim Y \kappa_c^2 W^2$. Deviations of row curvature from the preferred curvature result in bending strain $\kappa_c - \kappa_0$ and corresponding bending elastic energy density $\sim B (\kappa_c - \kappa_0)^2$. Hence, it can be expected that row bending elasticity dominates over lattice shearing in the narrow ribbon limit, such that rows keep natural curvature $\kappa_c \approx \kappa_0$.  In this regime, we therefore expect a power-law accumulation of elastic energy from the cost of shearing curved rows, $\epsilon_{\rm ex} (W \to 0) \sim Y \kappa_0^2 W^2$.  When the width grows sufficiently large, the shear cost to maintain preferred curvature $\sim Y \kappa_0^2 W^2$  eventually overwhelms the bending cost to {\it unbend} rows, $\approx B \kappa_0^2 /2$, and hence in this large width regime, we expect $\kappa_c \to 0$ and the elastic energy density saturates to the cost to unbend rows $\epsilon_{\rm ex} (W \to \infty) = B \kappa_0^2/2$.  We can use these energetic arguments to estimate that the crossover between the shear accumulation and flattening occurs at a characteristic {\it flattening size}

%Hence, when $W \to 0$ this shear cost vanishes, and it can be expected that row bending elasticity dominates over lattice strain, such that rows bend with their natural curvature $\kappa_c \approx \kappa_0$.  In this regime, we therefore expect a power-law accumulation of elastic energy from the cost of shearing curved rows, $\epsilon_{\rm ex} (W \to 0) \approx Y \kappa_0^2 W^2$.  When the width grows sufficiently large, the shear cost to maintain preferred curvature $\approx Y \kappa_0^2 W^2/2$  eventually overwhelms the bending cost to {\it unbend} rows, $\approx B \kappa_0^2/2$, and hence in this large width regime, we can instead expect $\kappa_c \to 0$.  Hence, in this large width regime, we then expect the elastic energy density saturates to the constant cost to unbend rows $\epsilon_{\rm ex} (W \to \infty) = B \kappa_0^2/2$.  We can use these energetic arguments to estimate that the crossover between the cost of shear accumulation and flattening energy occurs at a characteristic {\it flattening size}
\begin{equation}
    W_{\rm flat} \approx \sqrt{\frac{B}{Y}} \propto \lambda ,
\end{equation}
such that for $w \ll W_{\rm flat}$ we expect powerlaw growth of the excess energy with increasing width due to accumulating shears, which eventually saturates to the flattening energy $\epsilon_\infty = B \kappa_0^2/2$ for $w \gg W_{\rm flat}$.

We note that this energetic argument oversimplifies the spatial dependence of elastic ground states. In particular, while the interior of wide ribbons flattens, there is always a bent and sheared zone, of width roughly equal to $\lambda$, near to the free edges.  The effect of this finite-size boundary layer is reflected in the dependence of excess energy in eq. (\ref{eq:excess}) on the ratio $W/\lambda$ which crosses over from quadratic dependence for $W \ll \lambda$ to asymptotically saturated for $W \gg \lambda$. Nonetheless, the simple argument captures the narrow ribbon scaling of the elastic energy accumulation and the fact that the crossover lengthscale is set by the characteristic ratio between bending and shear moduli, $\lambda$.  In figure \ref{fig:width}(f), we analyze the elastic excess energy as function of ribbon width, and show that by rescaling the $\epsilon_{\rm ex}$ by the flattening energy $\epsilon_\infty$, and the width by $\lambda$, collapses the width dependence of excess energy for different values of row curvature (i.e. different $\phi$ values of WJ particle edge taper).  

This analysis and the continuum elastic energy show that the {\it range} accumulation versus saturation of excess energy is controlled by the characteristic elastic length $\lambda$, which is itself determined by the ratio of row bending to inter-row shear stiffness, $B/Y$.  The discrete-to-continuum mapping for WJ particle assemblies, eqs. (\ref{eq:Y}) and (\ref{eq:B}), shows that both elastic moduli are controlled by the ratio of the interaction depth to the square of the interaction range, $u_0/r_c^2$, as the elastic strain between rigid particles can only be born by deformation of those interparticle bonds.  However, these two elastic constants depend differently on the repulsive interactions in the binding regions, the relative strength of which is parameterized by $R$.  Heuristically, this can be understood via the pair-wise elastic modes shown in~figure \ref{fig:pair}(b).  Stretching or shearing two properly oriented, bond WJ particles simple loads all through binding sites in parallel, such that this modulus is simply set by the difference between stiffness of attractive and repulsive ``springs", i.e. $Y = 2 k_{\rm a} - k_{\rm r} \propto (1-R)$.  In contrast, bending the angle between bound particles simply pivots around the central repulsive site, loading only the attractive ``springs" under compression/tension, such that $B k_{\rm a} t^2/2$, is independent of $R$.  Hence, the elastic length that controls shape flattening in WJ particle assemblies varies with repulsive strength as 
\begin{equation}
\label{eq:lambdaR}
    \lambda = \frac{t}{\sqrt{(1- R)} } .
\end{equation}
In the attractive-only case ($R=0$) this length scale is limited by the discrete size of the particle because $t \leq d$.  However, as shown in~figure \ref{fig:pair}(c), as $R$ increases the effective elastic cost of shear and stretch deformations of the 2D lattice decreases, while the elastic cost of bend deformations is unchanged, leading to an increase of the elastic scale $\lambda$, and ultimately a divergence as $R \to 1$.

\begin{figure*}
    \centering
    \includegraphics[width=0.9\linewidth,height=\textheight,keepaspectratio]{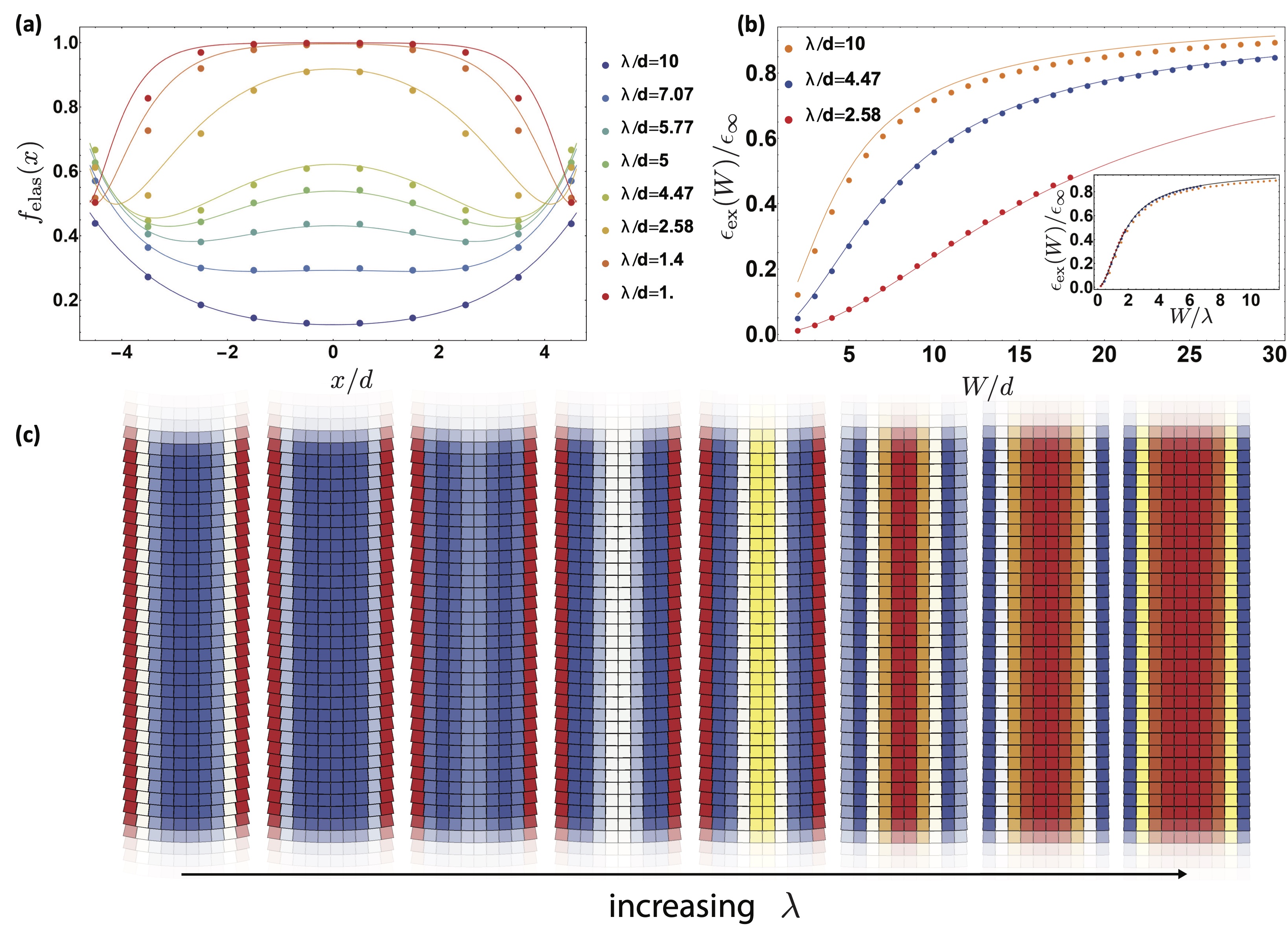}
    \caption{(a) The elastic energy density of infinite ribbons with $W/d = 10$ and varying $\lambda$ via varying $R=0-0.99$. The ribbons with values of $\lambda$ larger than the ribbon width have not yet begun the flattening transition here, while the ribbons with $\lambda$ less than the ribbon width have begun the flattening transition. %This gives evidence that the flattening transition is controlled by $\lambda$, i.e. the strength of the repulsive site relative to the cohesive energy. 
    The data points are from the numerical energy minimization, and the solid lines are from the continuum elasticity theory. (b) Another way of understanding this idea is in the rate of approach of the mean elastic energy density to its asymptotic value, $\epsilon_\infty$. Inset shows that the curves collapse upon rescaling the assembly width by $\lambda$. (c) Infinite ribbons, $W/d = 10$ and $\lambda/d = 1 - 10$ colored by excess interaction energy.}
    \label{fig:lambda}
\end{figure*}

In figure \ref{fig:lambda}(a) we test the dependence of the excess elastic energy in WJ particle ribbons on $\lambda$, via its dependence on relative repulsion/attraction strength between neighbors.  Here, we compare ribbons at a fixed width of $W = 10d$ but with several values of $R = 0 $ to 0.99 (corresponding to a predicted range of $\lambda/ d \simeq 1$ to 10).  Comparison of the spatial profiles of elastic energy density $f_{\rm elas}(x)$ shows a transition from the shape-flattened energy profile for $W/\lambda \gg 1$, which is largest and concentrated in the interior of the ribbon, to uniform bending profile for $W/\lambda \approx 1$, which is distributed throughout the assembly, but largest in the high-shear zones near the free boundaries.  The continuum theory shows agreement with  discrete-particle ground states for the magnitudes and qualitative spatial patterns of elastic distortion over the full range of repulsion strengths, and shows full quantitative agreement in the regime where the elastic scale extends much larger than the particle size, i.e. $\lambda \geq 5 d$, when row curvature (and its elastic effects) are distributed throughout the assembly.  

The effect of the repulsive strength on the range of frustration propagation is further reflected in the dependence of excess energy on ribbon width, compared in~figure \ref{fig:lambda}(b), for three values of repulsive strength.  These show that all elastic ground states follow the same basic trends of power-law accumulation at small widths followed by a saturation to a finite (flattening) cost a large size.  However, we find that increasing the strength of the repulsive pivot can have a profound effect on the {\it range} over which that accumulation occurs.  For more modest repulsion, when $\lambda = 2.6 d$, $\epsilon_{\rm ex}(W)$ very quickly reaches the saturating regime for $W\geq 5 d$, consistent with the fact that shear is confined to effectively narrow regions beyond that size.  However, for large repulsive strength, when $\lambda = 10 d$, the shear-accumulation deformation extends across the entire assembly until $W$ greatly exceeds this multi-particle dimension, ultimately driving a transition to the shape-flattened configuration only at especially large sizes. 

In the following section, we consider the effect of this variable range of accumulating vs. saturating elastic cost of frustration, and its dependence on the repulsion/attraction strength ratio $R$, on the thermodynamics of finite ribbon width selection.

\section{Thermodynamics of self-limited ribbon widths}

\label{sec: thermo}

Here we describe the thermodynamic competition between cohesion, which favors bulk structures, and frustration, whose elastic costs grow domain size, and its effects on the equilibrium width of ribbon domains of WJ particles.  We consider the infinite length, $H \to 0$, limit of the generic model introduced in eq. (\ref{eq: epsilon}), yielding
\begin{equation}
\label{eq: epsribbon}
    \epsilon_{\rm ribbon} (W) = \epsilon_0 + \frac{2 \Sigma}{W} + \epsilon_{\rm ex} (W) .
\end{equation}
In the WJ model, the bulk cohesive energy density (i.e. modulo the elastic costs of frustration) is $\epsilon_0= - 4u_0(1-R)/d^2 $ while the energy per unit length due to the fewer cohesive bonds is $\Sigma = 2u_0(1-R)/d$, while the excess cost of frustration captured by $\epsilon_{\rm ex} (W)$.  Assuming that thermodynamic ground state maintains cohesion throughout the assembly, and neglecting the possibility of internal defects, the width-dependence of excess energy is modeled by the continuum solutions in eq. (\ref{eq:excess}).  For a system in the canonical ensemble, well above the point of aggregation (i.e. supersaturated conditions), we consider a thermodynamic model where all (but a negligible fraction) of WJ particles assembled into ribbon domains, which we assume to have a uniform width $W_*$, momentarily neglecting fluctuations of width.  In this case, the width of equilibrium WJ ribbons is determined by the minimization of $\epsilon_{\rm ribbon} (W)$ eq. (\ref{eq: epsribbon})~\cite{hagan_equilibrium_2020}, yielding an equation of state $\Sigma(W_*)$ relating the edge energy to equilibrium width $W_*$
\begin{equation}
\label{eq: eos}
    \Sigma(W_*) \equiv \frac{W_*^2}{2} \epsilon_{\rm ex}' (W_*) = \frac{\epsilon_\infty  \lambda}{2} \Big[\tanh (W/\lambda) - \frac{\lambda}{W}\sech^2 (W/\lambda)  \Big] ,
\end{equation}
where $\epsilon_\infty = B \kappa_0^2/2$ is the elastic cost of {\it flattening} the preferred row curvature in the limit of large widths.  The variation of $W_*$ with $\Sigma$ encodes the intuitive result that increasing edge energy relative to elastic costs of deformation favors increasing the equilibrium width of assembly.

\begin{figure*}
    \centering
    \includegraphics[width=0.9\linewidth,height=\textheight,keepaspectratio]{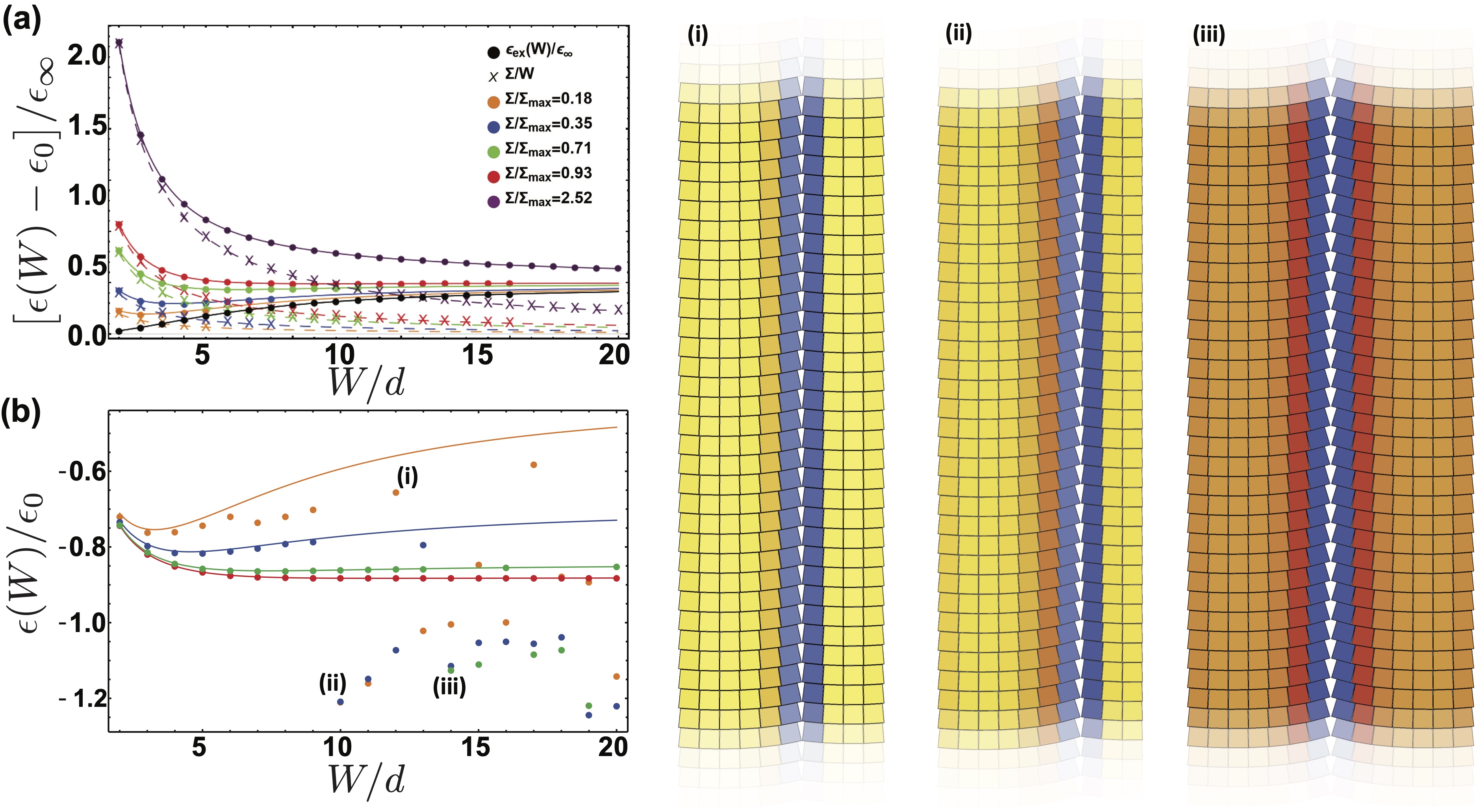}
    \caption{(a) The total energy per WJ particle in ribbon assemblies (minus bulk contribution) filled circles and solid curves, and the contributions due to surface energy density, 'X' symbols and dashed curves, and elastic deformations $\epsilon_{ex}$, shown in black, for varying $r_c/d$ and $R=0.95$. With increasing surface energy, the optimal ribbon width grows from $W=2d$ for $\Sigma/\Sigma_{max} = 0.18$ to $W=5d$ for $\Sigma/\Sigma_{max} = 0.35$ and then there is no finite minimum for larger $\Sigma/\Sigma_{max}$. (b) The corresponding total free energy density, for the same parameters and showing highly deformed, low-energy structures at larger $W/d$. (c) Typical low-energy structures that deviate from theory show internal boundaries (i.e. weakly cohesive cracks between sub-ribbons).}
    \label{fig: epsilonW}
\end{figure*}

The predicted dependence of selected width on edge energy follows from simple consideration of the small- and large-$W$ thermodynamics.  For small-$W$ thermodynamics, the excess energy grows quadratically due to width-dependent shearing between curved rows $\epsilon_{\rm ex} (W \ll \lambda) \approx Y \kappa_0^2 W^2$, which balances the edge cost $2 \Sigma/W$ at a width that grows as fractional power with $\Sigma$ and decreases with preferred row curvature as $W_* \approx (\Sigma/Y)^{1/3} \kappa_0^{-2/3}$.  Hence, the equilibrium size grows with the ratio of edge energy (i.e. cohesion) to elastic moduli.  Both $\Sigma$ and $Y$ are proportional to the net depth in interparticle binding (i.e. $2u_0(1-R)$), but $Y$ is also inversely proportional to the square of the interaction range $r_c$, hence the ratio $\Sigma/Y \propto  r_c^{2}$ decreases as interactions become shorter ranged, and effectively stiffer.  

In figure \ref{fig: epsilonW} we plot the (normalized) total energy density vs. ribbon width for varying $r_c/d = 0.181, 0.256, 0.363, 0.417$ corresponding to varying $\Sigma/\Sigma_{max}$ defined below.
%WJ particles with fixed shape frustration (i.e. constant $\phi$) and but for variable interaction range $r_c$, from $1.12\sigma$ to $2.58\sigma$.  
These results %confirm the basic result 
illustrate that for short enough $r_c$, where cohesion to elastic stiffness $\Sigma/Y\propto r_c^2$ is also small, that there is a minimum in $\epsilon_{\rm ribbon} (W)$ at a finite $W_*$, the value of which grows with interaction range.  We note, however, beyond a certain value of interaction range there is no minimum in the range up to 20 particles in width, the upper limit of discrete particle assembly sizes studied.  The equation of state from continuum elasticity theory indeed predicts that there is an upper limit to range of cohesive energies where a finite-$W_*$ minimum exists.  This follows from taking the infinite width limit of eq. (\ref{eq: eos})
\begin{equation}
    \lim_{W_* \to \infty} \Sigma(W_*) \equiv \Sigma_{\rm max} = \frac{\epsilon_\infty  \lambda}{2} ,
\end{equation}
which implies that equilibrium width {\it diverges} in the limit of $\Sigma \to \Sigma_{\rm max}$.  Specifically, eq. (\ref{eq: eos}) predicts a continuous divergence of the form $W_*/\lambda \sim - \log \sqrt{1-\Sigma/\Sigma_{\rm max}}$ in the asymptotic limit $W_* \gg \lambda$. The upper critical value of $\Sigma$ for finite ribbon width is a consequence of the shape-flattening transition.  When ribbons exceed the size $W_{\rm flat} \approx \lambda$, it is favorable from an elastic point of view to expel row curvature to a finite boundary layer at the free edge of the ribbon.  Notably, relative to the flattened interior, the row curvature within the boundary layer relaxes the elastic energy density by order $\epsilon_\infty = B \kappa_0^2/2$ over a range $\lambda$.  Hence, in this large $W \gg \lambda$ regime, elastic relaxation at boundary layers effectively decreases the edge energy by $- \lambda \epsilon_\infty/2=-\Sigma_{\rm max}$.  This implies that for sufficiently weak cohesion, when $\Sigma<\Sigma_{\rm max}$, the elastically renormalized edge energy of ribbons is {\it negative}, in effect stabilizing finite ribbon widths (in combination with stabilizing sub-leading terms in the elastic energy). When $\Sigma > \Sigma_{\rm max}$, it becomes favorable for the assembly pay the bulk elastic costs to expel frustration everywhere, and grow ribbons to infinite width.  

\begin{figure}[h]
    \centering
    \includegraphics[width=0.95\linewidth,height=\textheight,keepaspectratio]{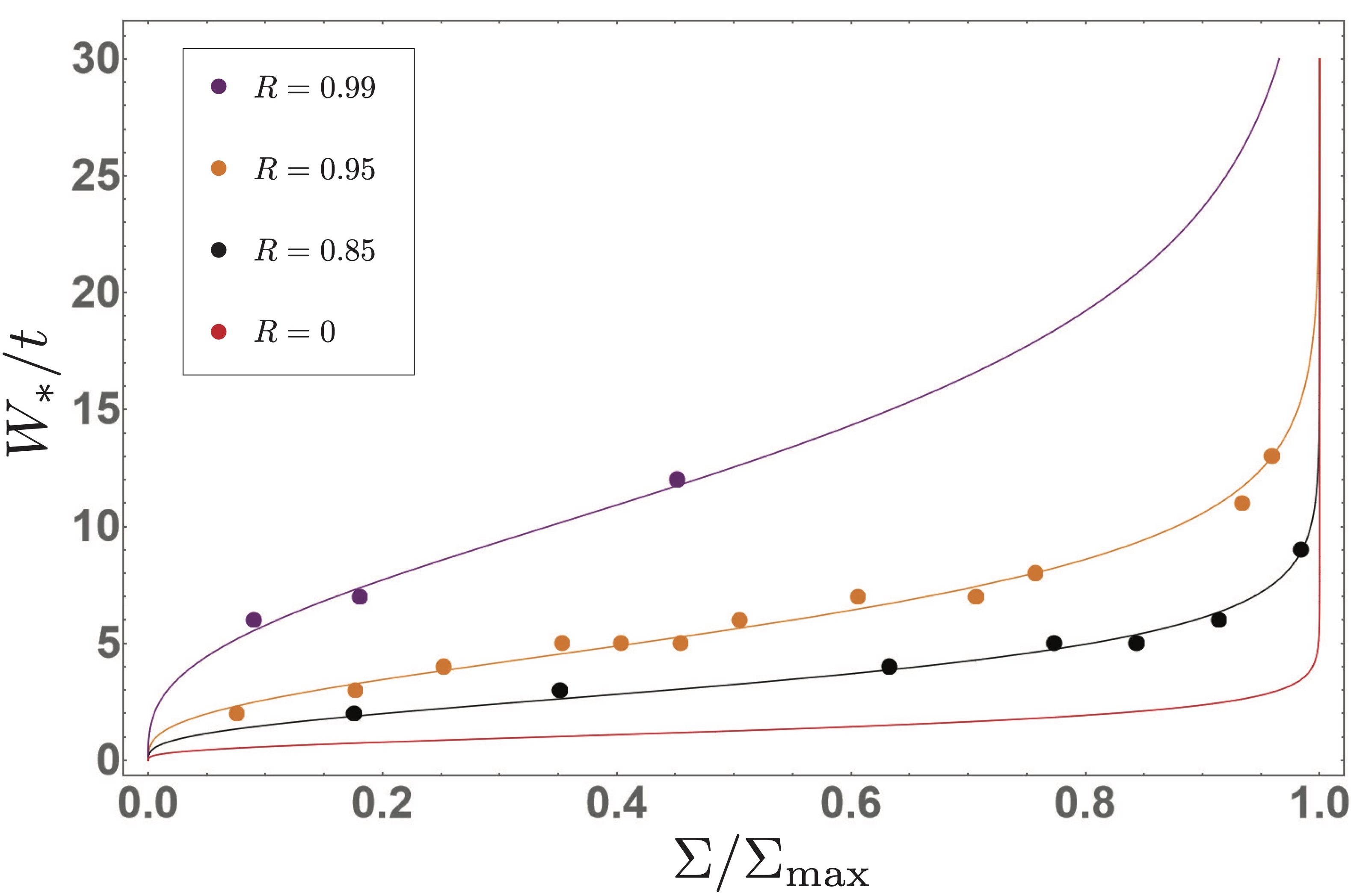}
    \caption{Equilibrium width of WJ ribbons versus surface tension for differing repulsive site strengths $R$, i.e. varying $\lambda$. As $R$ gets larger, the range of target ribbon widths before the divergent growth increases. The data points are from the numerical energy minimization, and the solid lines are from the continuum prediction, eq. (\ref{eq: eos}).}
    \label{fig: WvsSigma}
\end{figure}

In figure \ref{fig: WvsSigma} we show the full predicted variation of equilibrium width $W_*$ as a function of $\Sigma$, comparing minima obtained from numerical ground states of the discrete WJ particle model to the continuum predictions.  In particular, we compare the equilibrium widths for variable ratios $R$ of repulsive to attractive strength of WJ bonds.  These show that, for a given value of $\Sigma/\Sigma_{\rm max}$, the equilibrium width is an increasing function of $R$.  This can be attributed to the effect of repulsive interactions to extend the range of frustration propagation before shape flattening (i.e. $W_{\rm flat} \approx \lambda \propto (1-R)^{-1/2}$), which effectively extends the size scales for the self-limiting assembly by reducing the elastic costs of row shear while maintaining stiffness of row curvatures.  

Notably, in the limit of pure attraction ($R=0$), the predicted finite widths from the continuum model do not exceed even a single particle over most of the range of $0<\Sigma<\Sigma_{\rm max}$ and we are not able to resolve an energy minimum with the discrete WJ calculations.  This demonstrates that, for this class of frustrated assembly, it is essential to carefully control the interparticle mechanics through the their microscopic design in order to have self-limitation that occurs at non-trivial (i.e. $W_* > d$) sizes.  In the next section, we consider {\it fluctuations} of ribbon width and their impact on the possibilities for self-limitation near to the {\it flattening transition} where continuum theory predicts a divergent finite width.   

We note briefly the breakdown of the continuum theory to capture the energetic ground states of WJ ribbons at large width, reflected in the energy density plots in figure \ref{fig: epsilonW}(b). Specifically, we find that discrete particle ground states eventually ``crack'' and form structures with energies that fall below the continuum theory, which assumes that structures maintain cohesive elastic contact at all sizes.  As the cohesive interactions in the discrete model have a finite range, they exhibit yielding beyond a critical bond strain.  

Examples of the lower-energy, ``cracked'' states found via numerical minimization are shown in figure \ref{fig: epsilonW}(c).  These take the form of weakly adhered states of two or more elastically coherent ``sub-ribbons''.  The existence of these lower-energy states, aggregates of finite-domains, is a generic consequence of $T=0$ considerations of frustrated assembly.  If there there is local minimum in energy density $\epsilon(W_*)$ at finite width $W_*$, then it is possible to construct an equal energy density state with twice the size (i.e. with $2 W_*/d$ particles in lateral dimension) from two separated finite ribbons.  Bringing these two sub-ribbons into at least weak cohesive contact can only lower the total energy, such that it is always possible to find ``multi-ribbon'' aggregates where $\epsilon(nW_*)<\epsilon(W_*)$ for any integer $n$.  This basic argument is consistent with the structure of cracked and weakly-adhered sub-ribbons resulting from the energy minimization, although at present, we make no attempt to rigorously identity the ground states of these post-yield morphologies.  Analysis of the onset of  yielding of elastically coherent ribbons shows that it occurs at larger sizes than the local minima $\epsilon(W_*)$ set by the competition between elastic costs of frustration and cohesion.  While strictly-speaking ground states of any particulate model of GFA are likely of the form of weak-aggregates of finite domains, it can be expected that at large enough temperature (but well above the critical aggregation concentration), entropic effects break such weak contacts and stabilize at states dominated by self-limited domains size $W \simeq W_*$.

\section{Width-fluctuations of self-limited ribbons}

In the above, we showed that energetic ground states of the WJ-particle model exhibit a selected, finite width below a critical upper limit of edge energy $\Sigma_{\rm max}$.  The continuum theory and discrete-particle ground states demonstrate that the size of the selected width is an increasing function of the ratio of the row bending to inter-row shear moduli, as well as $\Sigma/\Sigma_{\rm max}$.  The continuum theory predicts that the growth of the selected width exhibits a second-order transition at the critical surface energy, that is, it diverges continuously to the bulk state as $\Sigma \to \Sigma_{\rm max}$.  This prediction seemingly implies that self-limited domains of arbitrarily large size are possible, provided that there is very careful control of the (subcritical) surface energy.  In this section, we describe a simple model of capillary fluctuations of the ribbon width, to understand how changes in the curvature of the minimum of $\epsilon(W_*)$ effect the possible range of controlled domain size.

We consider an infinite length ribbon with a mean finite width $W_*$, but with variable width $W(z) = W_* +\delta w(z)$ as a function of vertical height in ribbon.  As detailed in~\ref{sec: fluct}, this leads to a change in ribbon free energy,
\begin{equation}
    \Delta F \big[ \delta W(z) \big] \simeq \frac{1}{2} \int dz \Big[\frac{\Sigma}{2} (\partial_z \delta W)^2 + M(W_*) (\delta W)^2 \Big] ,
\end{equation}
where the first term derives from the capillary cost of excess edge length, while the second term describes the harmonic cost of deviations from the frustration selected width with,
\begin{equation}
\label{eq: M(w)}
    M(W_*) \equiv \partial_W^2 \big[ W \epsilon(W) \big] \Big|_{W_*}= \frac{ 2 \epsilon_{\infty}}{\lambda} \sech^2 \Big( \frac{W_*}{\lambda} \Big) \tanh \Big( \frac{W_*}{\lambda} \Big) .
\end{equation}
Note that as $W_*\gg \lambda$,  $M(W_*)$ decreases exponentially to zero, which indicates that as $W_*$ grows large, the minimum becomes more and more shallow as a consequence of the shape-flattening of large domains.  Hence, as the mean width grows arbitrarily larger than $\lambda$, so too should the thermal fluctuations of width.  Considering capillary width modes $\delta W(k) = \int dz ~ e^{ikz} \delta W(z)$ of wavevector $k$, equipartition at finite $T$ gives $\langle | \delta W(k)|^2 \rangle = k_B T /\big[\Sigma k^2/2 + M(W_*) \big]$, or at a given local height along the ribbon
\begin{equation}
\label{eq: deltaW}
    \langle |\delta W(z)|^2 \rangle = \frac{k_B T}{\sqrt{2 \Sigma M(W_*) } } .
\end{equation}
This result suggests that absolute thermal width fluctuations diverge both as edge energy vanishes ($\Sigma\to 0$) and as ribbons approach the bulk state ($\Sigma\to \Sigma_{\rm max}$) where $W_* \to \infty$.  

\begin{figure}
    \centering
    \includegraphics[width=0.95\linewidth,height=\textheight,keepaspectratio]{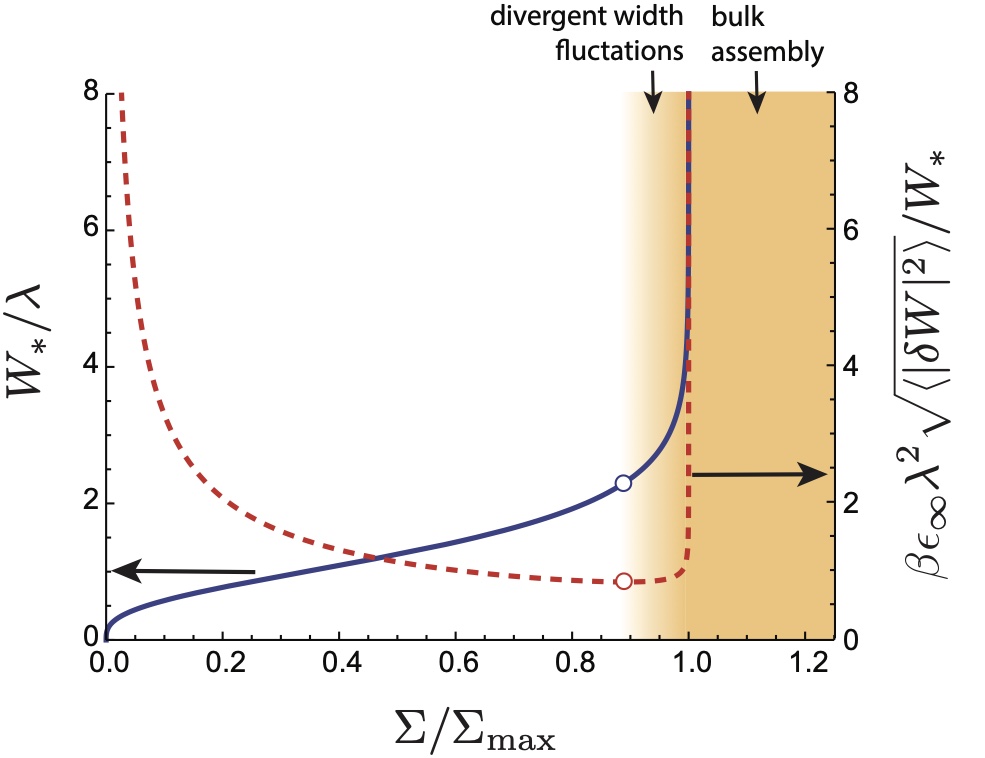}
    \caption{Analysis of thermal capillary fluctuations of ribbon width as a function of scaled edge energy.  The solid blue curve shows the predicted mean width $W_*$ of the ribbon (left axis) and the dashed red curve shows the ratio of r.m.s. width fluctuations to mean width (right axis). Shaded orange denotes the region of increasing minimal relative functions, which rapidly diverge as $\Sigma  \to \Sigma_{\rm max}$.  }
    \label{fig: fluctuations}
\end{figure}

In figure \ref{fig: fluctuations}, we plot the relative magnitude of width fluctuations as a function of edge energy, and compare these to the growth of mean width $W_*$ over the same range.  We note that the relative width fluctuations depend non-monotonically on surface energy, and in particular, they grow unbounded at the transition to the bulk state.  Hence, while the mean finite width extends continuously to infinite size, this point is preempted by divergent width fluctuations. As these relative fluctuations always diverge close to the (mean-field) bulk transition, we interpret the practical range of self-limitation as delimited by the growth of size fluctuations.  For simplicity, we can estimate the range of size-controlled, width-limited assembly as the range of $\Sigma$ for which relative fluctuations are minimal (i.e. magnitude of $\langle |\delta W|^2 \rangle^{1/2}/W_*$ rapidly diverges beyond this point).  From eq. (\ref{eq: deltaW}) we estimate that self-limitation is occurs only for $\Sigma < 0.9 \Sigma_{\rm max}$.  This corresponds to a maximal size of the self-limited state $W_{\rm max} \simeq W_*(0.9 \Sigma_{\rm max}) \simeq 2.5 \lambda.$

\section{Concluding remarks}

In summary, we have introduced a theoretical model for a new class of GFA: WJ particle assemblies.  The relatively simple geometric design of WJ particles makes it possible to trace microscopic features of the WJ subunits (i.e. their shape and interactions) to the emergent elasticity of inter-particle misfit gradients and their thermodynamic consequences for self-limitation of ribbon widths. Notably, this mapping from particulate properties to mesoscale description demonstrates that particle interactions and elasticity of the assembly cannot be considered independently for GFA, particularly for the case where subunit shapes are relatively rigid in comparison to the inter-particle bonds.  

We find that frustration propagation in WJ ribbons relies on the competition between two elastic modes: row-(un)bending elasticity, dominant for narrow widths; and inter-row shear elasticity, which dominates at large widths and forces ribbon interiors to flatten. The crossover between these two regimes determines an elastic {\it flattening} scale $W_{\rm flat} \approx \lambda$, controlled only by the stiffness ratios of these two elastic modes, and notably independent of degree of shape frustration (i.e. preferred row curvature $\kappa_0$).  Additionally, this length scale defines the boundary between regimes of accumulating ($W \ll \lambda$) and asymptotically saturating ($W \gg \lambda$) elastic costs.  As pointed out previously~\cite{hagan_equilibrium_2020,Meiri2021}, generic considerations of GFA thermodynamics suggest that self-limiting domain formation relies on the existence of an accumulating regime of elastic costs, while the asymptotic saturation of shape-flattening progression of elastic ground states typically implies a thermodynamic escape to infinite size beyond a critical surface cohesion.  We show for the WJ model at $T=0$, that this manifests for planar and vertically oriented ribbons as a second-order transition between finite width and unlimited (bulk) domains at a critical edge energy $\Sigma_{\rm max}$ which itself is proportional to the product of the flattening elastic cost and the flattening size. Consideration of the finite-temperature fluctuations show that the depth of the energy minimum that selects mean ribbon width becomes arbitrarily shallow as this (mean-field) critical point is approached. Analysis of these divergent width fluctuations suggests that practical regime of self limitation is rather better described by the regime $W_* < 2.5 \lambda$.  Hence, both of these results imply that the range of accessible self-limiting states, their size and cohesive energy, is strongly dependent the elastic length scale defined by the ratio of bending to shear stiffness.  In light of this, we demonstrate how spatial design of locally attractive and repulsive patches in the binding domain can effectively extend this characteristic size, and thereby the effective range of self-limitation, to size scales that extend for multiple particles.

We note that the relatively simple design of WJ particles has clear potential for implementation in experimental studies of intentionally shape-misfitting particle assembles.  In particular, recent approaches to DNA origami particles realize anisotropic, self-assembly particles with anisotropic shapes that defined the relative geometries of bound neighbors~\cite{berengut2020self}, but also combine specific (i.e. lock and key type) interactions between specific binding sites~\cite{Gerling2015, Sigl2021, Hayakawa2022}, of the type incorporated in the WJ model.  Our results hold key lessons for the experimental design of such particle.  Beyond the intuitive result that the maximal cohesive strength for limitation grows with particle taper (squared), we find that the practical range of accessible self-limited sizes is dominantly controlled by the ratio of elastic modes of inter-particle deformation.  Hence, realizing experimental WJ particles that exhibit non-trivial ranges of self-limiting assembly (i.e. larger than $\sim 1-2$ particles wide) requires careful engineering of interactions that selectively control distinct deformations (i.e. bending vs. shear/stretch).  Here, we find that this requires careful control of interactions for particles of strictly rigid shape.  In experimental systems, such as DNA origami particles, it might also be expected that intra-particle deformations may contribute to (and potentially dominate in certain situations) the elastic contributions of interaction~\cite{Tyukodi2022}. Although as of yet, a sufficiently careful analysis of the relative compliance of such particles vs. their binding sites remains to be conducted for DNA programmed shape-defined particles.

We conclude by remarking on two possible extensions of the WJ model, both of which will be addressed in future study. First, we note that it is straightforward to consider more anisotropic binding properties of WJ particles (i.e. strength and stiffness of horizontal vs. vertical row bonds need not be the same).  In the case of anisotropic binding, vertical and horizontal edge energies become unequal.  When horizontal edge energy is sufficiently large compare to vertical edge energy, it is possible to shift the thermodynamic ground state from vertical type ribbons to horizontal and globally curved ribbons, of the type shown in~figure \ref{fig:vertical_vs_horizontal}(b).  Hence, the general phase diagram of even strictly planar assembly of WJ particles exhibits polymorphism between different states of finite thickness ribbons.  It remains to be seen if the frustration build up in this distinct class of curved ribbon morphologies exhibits distinct thermodynamic growth of self-limiting dimensions, and whether that growth relies on a distinct set of elastic modes (and corresponding moduli).  

Beyond the polymorphism of planar assembly, it is far from clear how ground states of the WJ model behave when allowed to buckle out of the 2D plane.  In certain 2D GFA models, such as assembly of pentagonal or heptagonal particles~\cite{Lenz2017}, it is possible to identify non-Euclidean geometries that fully relax frustration that would be required by planar assembly (e.g. spherical tilings of pentagonal particles).  In this context, it reasonable to ask if there are surface shapes that relax all, or at least some of the frustration of WJ particle assembly, and if so, how does this modify thermodynamics of domain formation?  One clue to resolving this question may be to note a similarity to aspects of WJ assembly and 2D textures of so-called ``bent core'' liquid crystals~\cite{Selinger2022, Meiri2022}.  Like the horizontal rows of WJ assemblies, such bent core mesogens favor uniform bending along their long axis, but without splaying the distances between those field lines.  Niv and Efrati derived generic compatibility conditions for such textures embedded in 2D surfaces of arbtirary shape, and showed that it is possible to achieve this uniform bend/zero splay texture on surfaces of constant, negative Gaussian curvature, thereby fully relaxing the frustration cost of orientational gradients in the liquid crystal~\cite{Niv2018}.  Therefore, we might expect at least some degree of frustration relaxation in WJ-particle assemblies if their membranous assemblies are sufficiently flexible to adopt such shapes.  However, it should be noted that WJ assemblies possess additional elastic constants beyond the orientational elasticity of 2D liquid crystals, such as inter-row shear and intra-row stretch moduli.  It is not yet known what are the complex 3D shapes that are favored by frustrated assembly WJ particles, how much of the elastic cost of planar assembly can be eliminated by out-of-plane deformation, and ultimately, how this alters the basic limits of self-limited domain formation relative to the planar case.

\begin{acknowledgements}
The authors are grateful to M. Wang, C. Santangelo and M. Hagan for valuable discussions on this work.  This research was supported by the US National Science Foundation through award NSF DMR-2028885, as well as through the Brandeis MRSEC on Bioinspired Materials NSF DMR-2011846.  Computational studies of WJ-particle ground states were performed on the UMass Cluster at the  Massachusetts Green High Performance Computing Center.
\end{acknowledgements}

\appendix

\section{Fluctuations of ribbon width}

\label{sec: fluct}

Here, we describe a continuum model for thermal width fluctuation of long-ribbons.  As we are interested in the limit where ribbons grow unbounded in length, we consider a case where $L \to \infty$.  We begin by decomposing the capillary modes of finite width ribbons into two modes.  Describing the a vertically oriented ribbon by the horizontal positions of the left and right edges as a function of vertical position $z$, $x_-(z)$ and $x_+(z)$, respectively.  I.e. for the ground state ribbon, we have that $x_+(z) = W_* + x_-(z) = {\rm const.}$.  The lengths $\ell_{\pm}$ of the free edges are simply,
\begin{equation}
   \ell_{\pm} = \int dz \sqrt{1 + (\partial_z x_{\pm})^2 } \simeq L + \frac{1}{2} \int dz ~ (\partial_z x_{\pm})^2 .
\end{equation}
The total edge energy of the ribbon is
\begin{multline}
    E_{edge} \simeq 2 \Sigma L +\frac{\Sigma}{2} \int dz ~\big[ (\partial_z x_+)^2+(\partial_z x_-)^2\big] \\ = 2 \Sigma L + \Sigma \int dz ~\big[ \frac{1}{4}(\partial_z W)^2  + (\partial_z \bar{x})^2 \big]
\end{multline}
where $W(z)= x_+(z) - x_-(z)$ is the local width and $\bar{x}(z) = \big[x_+(z) - x_-(z)\big]/2 $ is the central axis of the ribbon.  We note that the second term in integrand, proportional to $(\partial_z \bar{x})^2$, penalizes tilting ribbon orientations away from the (low edge energy) bond vertical, but it is decoupled from fluctuations of width $W(z)$ to quadratic order.  Hence, below we consider only the edge energy penalty to width gradients.

In addition to penalties in edge energy, edge fluctuations are suppressed by combined effects of frustration and boundary energy that select a finite equilibrium width.  The free energy per unit of uniform width ribbon is simply
\begin{equation}
    F/ L= W \big[\epsilon (W) - \mu \big] \simeq W_* \big[\epsilon (W_*) - \mu \big] + \frac{(\delta W)^2}{2} \partial_W^2 \big[ W \epsilon (W) \big] \Big|_{W_*} ,
\end{equation}
where $\mu$ is the chemical potential cost of adding a free particle to a ribbon and $\delta W = W - W_*$ is the fluctuation from the selected width.
Using eq.~(\ref{eq: eos}) and defining
\begin{equation}
M(W_*) \equiv \partial_W^2 \big[ W \epsilon (W) \big] \Big|_{W_*} = \frac{ 2 \epsilon_\infty}{\lambda} \sech^2\Big( \frac{W_*}{\lambda} \Big)\tanh \Big( \frac{W_*}{\lambda} \Big)
\end{equation}
we have the free energy functional for width fluctuations,
\begin{eqnarray}
     \Delta F \big[\delta W(z) \big]\simeq \frac{1}{2} \int dz ~ \Big[\frac{\Sigma}{2} (\partial_z \delta W)^2 + M (\delta W)^2 \Big] \\   = \frac{1}{2} \int \frac{ d k }{2 \pi} \Big[\Sigma k^2/2 + M \Big] | \delta W (k) |^2,
\end{eqnarray}
where
\begin{equation}
    \delta W(k) = \int dz ~e^{ikz}  \delta W(z) ,
\end{equation}
is the Fourier-transformed width fluctuation.  From equipartition at finite temperature we have
\begin{equation}
    \langle |\delta W(k)|^2 \rangle = \frac{1}{\beta\big(\Sigma k^2/2 + M \big) }
\end{equation}
with $\beta^{-1} = k_B T$, which can be Fourier transformed back to yield the Gaussian thermal fluctuations of width,
\begin{equation}
\langle |\delta W(z)|^2 \rangle = \int \frac{d k}{2 \pi} \frac{\beta^{-1}}{\big(\Sigma k^2/2 + M \big) } = \frac{\beta^{-1}}{ \sqrt{2 \Sigma M }}.
\end{equation}

\section{WJ particle model parameters}

\label{sec: parameters}
As described in the main text, the geometry of the WJ particles are the same for all of our simulations. The dimensionless parameters that were changed are $R$, $r_c/d$, and $u_0/u_{\rm ex}$. The parameters used in each figure are given in the following tables.
\begin{center}
\begin{tabular}{ |c|c|c|c| } 
 \hline
  & $R$ & $r_c/d$ & $u_0/u_{\rm ex}$\\ 
 Fig. \ref{fig:assembly-example} & 0 & 0.323 & 1 \\ 
 Fig. \ref{fig:vertical_vs_horizontal} & 0 & 0.323 & 1 \\
 Fig. \ref{fig:width} & 0.95 & 0.417 & 1.85 \\ 
 Fig. \ref{fig:lambda} & 0--0.98 & 1.393 & 20 \\
 Fig. \ref{fig:lambda} & 0.99 & 1.358--1.461 & 19--22 \\
 Fig. \ref{fig:lambda} & 0.95 & 0.424 & 1.85 \\
 Fig. \ref{fig:lambda} & 0.85 & 0.194 & 1.4  \\
 Fig. \ref{fig: epsilonW} & 0.95 & 0.184--0.696 & 0.35--5 \\
 Fig. \ref{fig: WvsSigma} & 0.99 & 0.440--0.985 & 2--10 \\
 Fig. \ref{fig: WvsSigma} & 0.95 & 0.121--0.429 & 0.15--1.9 \\
 Fig. \ref{fig: WvsSigma} & 0.85 & 0.082--0.194 & 0.25--1.4 \\
 \hline
\end{tabular}
\end{center}

\setcounter{section}{1}

\section*{References}
\bibliographystyle{unsrt}
\bibliography{WJbib}

\end{document}